\documentclass[10pt]{article}

\usepackage{geometry}
\usepackage{amsmath}
\usepackage{latexsym}
\usepackage{amsfonts}
\usepackage{mathrsfs}
\usepackage{upgreek}
\usepackage{graphics}
\usepackage{graphicx}
\usepackage{epstopdf}
\usepackage{epsfig}
\usepackage{amsbsy}
\usepackage{amsfonts}
\usepackage{amsmath}
\usepackage{amssymb}
\usepackage{bm}
\usepackage{color}
\usepackage{setspace}

\def\bvp{\boldsymbol v^{\textrm{p}}}
\def\bvs{\boldsymbol v^{\textrm{s}}}
\def\bg{\boldsymbol g}
\def\bvf{\boldsymbol v^{\textrm{f}}}

\def\rhos{\rho^{\textrm{s}}}
\def\rhop{\rho^{\textrm{p}}}
\def\rhof{\rho^{\textrm{f}}}
\def\pp{p^{\textrm{p}}}
\def\pf{p^{\textrm{f}}}

\def\bff{\boldsymbol{f^{\textrm{pf}}}}
\def\be{\boldsymbol{e}}
\def\gammaf{\dot\gamma^{\textrm{f}}}
\def\gammap{\dot\gamma^{\textrm{p}}}
\def\gammas{\dot\gamma^{\textrm{s}}}

\newcommand\pd[2]{ \dfrac{\partial {#1}}{\partial {#2}}}

\title{Flow of suspensions in a hydraulic fracture consisting of Herschel-Bulkley fluid and spherical particles}

\author{E.V. Dontsov, S.A. Boronin, and A.A. Osiptsov}

\begin{document}
\maketitle
 
\abstract{\noindent The purpose of this study is to develop a model for the flow of suspensions consisting of Herschel-Bulkley fluid mixed with spherical particles. In particular, the focus is to investigate the effect of non-Newtonian rheology of the carrying fluid on the flow behavior of a suspension. Two-dimensional steady flow problem in a vertical channel is considered, in which both the pressure gradient and gravity drive the suspension flow. Dependence of the velocity profile and particle concentration across the channel on the fluid rheology parameters and orientation of the pressure gradient is investigated. It is found that the non-uniform particle distribution in the flow across the channel leads to the non-uniform density of the suspension, which causes sinkage of the denser regions and promotes downward migration of the particles even without slip velocity. Particle and suspension fluxes are calculated for various fluid rheologies and pressure gradient orientations. The effect of slip velocity between the phases is added via filtration term that captures fluid flow once particles reach the maximum concentration and stall, and via the settling term that describes gravitational particle settling.}
\vspace{3 mm}


\section{Introduction}

Appropriate placement of proppant particles is one of the key challenges that needs to be addressed in hydraulic fracturing operations~\cite{Econo2000,King2012}. It is therefore necessary to develop proppant transport models, which would help to address the challenge. This problem has been investigated by multiple authors in the past~\cite{Dane1978,Pear1994,Mobbs2001,Shokir2007,Eskin2008,Boro2010,Dont2014b,Lecam2014,Shioz2016,Donts2019,Isaev2023}. The most common and practical approach in proppant transport modeling, is to assume that the proppant is simply carried by fracturing fluid with the same velocity and there is a slip velocity due to gravity-induced settling~\cite{Adachi2007}. The proppant influences the slurry viscosity according to a given law, which makes the effective viscosity larger for increasing proppant concentrations. The viscosity becomes singular once the proppant concentration reaches a critical value and the slurry flow stops. This is a very practical approach, which essentially captures all zeroth order effects. However, there are multiple effects that are ignored, see e.g. a review paper~\cite{Osipts2017}. To better understand the influence and importance of the fluid flow effects, the aim of this study is to construct a proppant transport model from the solution of the suspension flow in a fracture.

It is important to consider a suspension rheology model, in which a transition to dense suspensions is captured. The reason for this lies in the fact that both dilute and packed suspensions can be present in a hydraulic fracture. The former typically occurs within the fracture, while the latter often happens in the vicinity of the fracture tip due to accumulation of proppant, for instance, due to either proppant settling or slurry dehydration caused by fluid leak-off into surrounding rock formation. Most of the existing models consider slurry as the mixture of a Newtonian fluid and spherical particles. Earlier approaches for modeling suspensions include the diffusive flux method~\cite{Leigh1987} and the suspension balance model~\cite{Nott1994}. In the diffusive flux approach, the particle concentration across the channel is modeled by a nonlinear diffusion equation, in which the particle flux is related to particle concentration and shear rate of the flow. The model relies on the expression for the flux, whose exact expression is based on experimental observations and features fitting parameters. This model was used by~\cite{Phil1992} to solve the problem of steady flow in a tube and in a channel. The model is able to capture shear migration, which leads to the nonuniform particle concentration, as well as the blunted velocity profile caused by such a particle distribution. Suspension balance model~\cite{Nott1994} is based on averaging of the microscale governing equations and can be sought as a generalization of the diffusive flux method. The model introduces particle pressure, which depends on concentration and shear rate. In this case, the particle migration is driven by the particle pressure gradient. This model also needs a set of closure relations, which can be only measured experimentally. It is also important to mention the study~\cite{Nott2011}, in which the suspension balance model is revisited. Yet, this theoretical model still needs input from experimental data.

In contrast to the diffusive flux method and the suspension balance model, more recent approaches characterize behavior of suspensions by the shear stress and normal stress (or particle pressure) that are induced by the applied shear rate $\dot\gamma$, namely
\begin{equation}\label{taupp}
\tau=\eta_s(\phi) \mu \dot\gamma,\qquad \pp=\eta_n(\phi) \mu\dot\gamma,
\end{equation}
where $\mu$ is the intrinsic fluid viscosity, $\eta_s(\phi)$ and $\eta_n(\phi)$ are the dimensionless effective shear and normal viscosities, and $\phi$ is the particle volume concentration. The functional forms $\eta_s(\phi)$ and $\eta_n(\phi)$ are typically measured experimentally in one form or another and are used to quantify the behavior of suspensions.

One of the models was proposed in~\cite{Morr1999}, in which shear and normal stresses of the slurry are given by
\begin{equation}\label{Morrmodel}
\eta_s =1+2.5\phi \Bigl(1-\dfrac{\phi}{\phi_m}\Bigr)^{-1} +K_s\dfrac{\phi^2}{(\phi_m-\phi)^2},\qquad \eta_n =K_n\dfrac{\phi^2}{(\phi_m-\phi)^2},
\end{equation}
where $K_s=0.1$ and $K_n=0.75$ are two numerical parameters. This model also satisfies the limit for dilute suspensions ($\phi\!\ll\!1$) $\eta_s=1+2.5\phi$~\cite{Eins1905}. For the non-shearing regions of the suspension, equations~(\ref{taupp}) and~(\ref{Morrmodel}) are combined to yield
\begin{equation}\label{Morrmodel2}
\tau\leqslant\dfrac{K_s}{K_n}\pp,\qquad \phi=\phi_m,
\end{equation}
which corresponds to the friction law with the friction coefficient that is equal to $K_s/K_n$. The rheological model~(\ref{Morrmodel}) and~(\ref{Morrmodel2}) was used in~\cite{Mill2006} to solve the problem of suspension flow in a channel. In particular, authors observed particle migration towards the center of the channel and blunted velocity profile, which agreed with the experimental observations. In addition, authors utilized the non-local stress calculation for evaluating the equilibrium particle concentration profile. The argument for doing this lies in the fact that particles ``feel'' the shear rate that is averaged over the distances on the order of the particle size. This non-local stress especially influenced behavior near the center of the channel, where the shear rate approaches zero.

Another model for suspension flow was proposed in~\cite{Boyer2011}. This model is based solely on the experimental results and can be summarized as
\begin{equation}\label{Boymodel}
\eta_s =1+2.5\phi \Bigl(1-\dfrac{\phi}{\phi_m}\Bigr)^{-1} +\Bigl[\mu_1+\dfrac{\mu_2-\mu_1}{1+I_0\phi^2/(\phi_m-\phi)^2} \Bigr]\dfrac{\phi^2}{(\phi_m-\phi)^2},\qquad \eta_n =\dfrac{\phi^2}{(\phi_m-\phi)^2},
\end{equation}
where $\mu_1=0.32$, $\mu_2=0.7$, and $I_0=0.005$ are the fitting parameters and $\phi_m=0.585$ is the observed maximum particle volume fraction. Behavior of the suspension for the maximum concentration can be deduced as
\begin{equation}\label{Boymodel2}
\tau\leqslant\mu_1\pp,\qquad \phi=\phi_m,
\end{equation}
where $\mu_1$ plays the role of a friction coefficient. As can be seen from comparing~(\ref{Morrmodel}) and~(\ref{Boymodel}), both models have similar expressions and behavior at small and high particle concentrations. At the same time, the models differ in terms quantitative predictions, which, however, can be handled by using suitable values for $K_s$ and $K_n$. 

The rheological model~(\ref{Boymodel}) was used in~\cite{Dont2014b} to solve the problem of a slurry flow between two parallel plates. In particular, two-dimensional fluid flow is considered and the effect of gravitational settling on the distribution of particles across the channel and slurry flow is investigated. It is shown that the presence of gravity leads to a fully two-dimensional flow profile, in which the direction of the slurry velocity changes across the channel width. However, in order to construct an amenable proppant transport model, the complex coupling with gravity was replaced by a simpler expression, in which the influence of gravitational settling on distribution of particle concentration across the channel was neglected. In this situation, the gravitational settling and the slurry flow problems become uncoupled and the settling is modeled by an additional vertical particle velocity, which, as per the assumption, does not influence distribution of particles across the channel. The particle concentration for the uncoupled problem (i.e. without gravity) features a zone of maximum particle concentration in the center of the channel, and the size of this zone is determined by the condition~(\ref{Boymodel2}). Once the shear stress exceeds the critical value, shear motion occurs and the particle concentration decays gradually towards the walls. This particle distribution also leads to a blunted velocity profile. In addition, the two-velocity model is utilized, in which the slip velocity between particles and fluid is included. Such a formulation led to the development of the slurry flow model that captures the transition from Poiseuille flow for low particle concentrations to Darcy's filtration law for high particle concentrations. This transition is not typically present in the analysis of the suspension flow since the shear rate is applied and the shear and normal stresses are measured. In contrast, the pressure gradient is applied in the problem of a channel flow. In this case, even if the particles reach the maximum concentration and are unable to move, the fluid is still able to flow through due to filtration. The developed proppant transport model was implemented for plane strain and pseudo-3D hydraulic fractures in~\cite{Dont2015,Dont2015b} and in~\cite{Shioz2016,Wang2018,Wang2018b,Yang2024} for more complex geometries of hydraulic fractures. 

The following modification to the rheology~(\ref{Boymodel}) was proposed in~\cite{Lecam2014}
\begin{equation}\label{Lmodel}
\eta_s =1+2.5\phi \Bigl(1-\dfrac{\phi}{\phi_m}\Bigr) -\dfrac{\phi^2}{\phi_m^2}+\Bigl[\mu_1+\dfrac{\phi_m}{\beta}\Bigl(1-\dfrac{\phi}{\phi_m}\Bigr) \Bigr]\dfrac{\phi^2}{(\phi_m-\phi)^2},\qquad \eta_n =\dfrac{\phi^2}{(\phi_m-\phi)^2},
\end{equation}
where $\mu_1=0.3$, $\beta=0.158$, and $\phi_m=0.585$. In addition, authors in~\cite{Lecam2014} suggested further compaction beyond the maximum value of $\phi_m$. In particular, equations~(\ref{taupp}) and~(\ref{Lmodel}) for the non-shearing suspension are rewritten as
\begin{equation}\label{Lmodel2}
\tau=\Bigl[\mu_1+\dfrac{\phi_m}{\beta}\Bigl(1-\dfrac{\phi}{\phi_m}\Bigr) \Bigr]\pp,\qquad \phi\geq \phi_m,
\end{equation}
which is used to calculate the particle volume fraction in the non-flowing regions. For instance, in the center of the channel flow, where the shear stress vanishes, the maximum volume fraction becomes $\phi_{rcp}=\phi_m+\mu_1 \beta\approx0.632$, which corresponds to the volume fraction of the random close packing. Authors in~\cite{Lecam2014} used the developed rheological model to solve the problem of a slurry flow between two parallel plates, which is extension of the work~\cite{Mill2006} for a different suspension rheology. The gravitational settling was not considered and the slip velocity between the particles and the fluid, which gave rise to the filtration term in~\cite{Dont2014b}, was neglected. Instead, authors focused on the problem of particle concentration evolution from the uniform profile at the inlet to the steady-state profile further away from it. One of the distinct features of the model~(\ref{Lmodel}) and~(\ref{Lmodel2}) is the ability to describe compaction of the particles beyond the value of $\phi_m=0.585$. As a result, the particle concentration experiences a linear growth from $\phi_m=0.585$ to $\phi_{rcp}=0.632$ within the non-yielding part of the suspension. 

It is worth mentioning the study~\cite{Oh2015}, in which flow of dense suspensions in a pipe was studied experimentally. Authors used the rheological model~(\ref{Lmodel}) with compaction~(\ref{Lmodel2}) and without compaction~(\ref{Boymodel2}). It is interesting to observe that both models failed to precisely capture the experimental results, albeit to a different degree. The predicted fluid velocity profiles were nearly identical for both models and agree with the experimental observations, while the particle concentration profiles were different. Results for two mean particle concentrations $\phi=0.52$ and $\phi=0.55$ indicate that there is a plateau with concentration $\phi_{rcp}$ in the center of the channel. The model with no compaction~(\ref{Boymodel2}) predicted the correct shape (i.e. constant concentration), but with the wrong value of $\phi_m$. On the other hand, the model with compaction~(\ref{Lmodel2}) predicted a gradual growth of the concentration from $\phi_m$ to $\phi_{rcp}$ instead of a simple plateau. The latter model arguably provides a somewhat better prediction, but still does not capture the experimental results and, most importantly, does not capture the behavior qualitatively in the center of the channel. Arguably, if one replaces $\phi_m$ with $\phi_{rcp}$ in the model~(\ref{Boymodel}) and~(\ref{Boymodel2}), then the experimental results for the pipe flow will likely be matched.

The study~\cite{Dago2015} extends the results of~\cite{Boyer2011}, namely the suspension rheology~(\ref{Boymodel}), to Herschel-Bulkley fluids. The shear and normal stresses are now computed in lieu of~(\ref{taupp}) as
\begin{equation}\label{tauppHB}
\tau=\dfrac{\eta_s(\phi)}{{\cal F}(\phi) } (\tau_0+k \dot\gamma^n {\cal F}(\phi)^n),\qquad \pp=\dfrac{\eta_n(\phi)}{{\cal F}(\phi) } (\tau_0+k \dot\gamma^n{\cal F}(\phi)^n),
\end{equation}
where $\tau_0$ is the yield stress, $k$ is the consistency index, and $n$ is the power-law exponent. The relation between~(\ref{tauppHB}) and~(\ref{taupp}) is based on the following considerations, see also~\cite{Chat2008}. It is assumed that the dynamics is determined by the local shear rates, that occur in between the particles. The relation between the local and global shear rates is taken in the form
\[
\dot\gamma_{local}={\cal F}(\phi)  \dot \gamma,
\]
where it is assumed that the proportionality function ${\cal F}(\phi)$ depends solely on the geometrical parameters and thus depends only on $\phi$. Then, the effective viscosity is computed based on the local shear rate and is substituted into~(\ref{taupp}) to obtain~(\ref{tauppHB}). Authors in~\cite{Dago2015} also determine that 
\begin{equation}\label{calF}
{\cal F}(\phi) =2\sqrt{\dfrac{\eta_s(\phi)}{1-\phi}},
\end{equation}
which is consistent with the analysis in~\cite{Chat2008}. They however note that ${\cal F}(\phi) =2.8\sqrt{\eta_s(\phi)}$ also yields a good agreement. Finally, the experiments in~\cite{Dago2015} focused solely on dense suspensions $\phi\gtrsim 0.45$ and the rheology~(\ref{Morrmodel}) with suitable parameters was used to fit the experimental data. In particular, authors used
\begin{equation}\label{Morrmodel3}
\eta_s =\beta_0 K_n+\alpha_0 K_n\dfrac{\phi}{\phi_m-\phi} +\mu_0 K_n\dfrac{\phi^2}{(\phi_m-\phi)^2},\qquad \eta_n =K_n\dfrac{\phi^2}{(\phi_m-\phi)^2},
\end{equation}
where $K_n=0.75$, $\beta_0=6$, $\alpha_0=4.6$, and $\mu_0=0.3$. One downside of the model~(\ref{Morrmodel3}) is that it does not reduce to the dilute regime for low concentrations~\cite{Eins1905}.

Rheology of suspensions that consist of Newtonian fluid with rigid cylindrical fibers is studied in~\cite{Tapia2017}. It is shown that the fibers introduce yield stress for both shear an normal stresses~(\ref{taupp}). These stresses vary with the aspect ratio and particle concentration. In addition, the maximum particle concentration $\phi_m$ was found to vary substantially with the aspect ratio, leading to smaller values of $\phi_m$ for longer particles. Once the yield stress and the maximum particle concentrations are accounted for, the values of $\eta_s(\phi)$ and $\eta_n(\phi)$ become universal. Interestingly, it is found that the normal and shear stresses~(\ref{taupp}) diverge as $(1-\phi/\phi_m)^{-0.9}$ near maximum concentrations, which is in contrast to $(1-\phi/\phi_m)^{-2}$ for spherical particles. At the same time, friction coefficient, which is equal to the ratio between the shear and normal viscosities is well approximated by the model~(\ref{Morrmodel}) near jamming and with suitable parameters.

As can be seen from the above literature review, there is a rich history of suspension flow modeling. For the purpose of this study, however, we focus on the effect of Herschel-Bulkley fluid rheology and aim to develop solutions for the flow in a vertical channel that are relevant for proppant transport. The paper is organized as follows. Section~\ref{secrheology} summarizes the adopted suspension rheology model. Section~\ref{secchannel} presents problem formulation for the flow in a channel, while Section~\ref{secchannelsol} presents solution for the problem. Then, Section~\ref{secfluxes} accommodates the effect of slip velocity and outlines expressions for fluxes, which is followed by discussion in Section~\ref{secdisc} and summary in Section~\ref{secsummary}.

\section{Adopted rheological model for slurry rheology}\label{secrheology}

\begin{figure}[h]
\centerline{\includegraphics[width=1.0\linewidth]{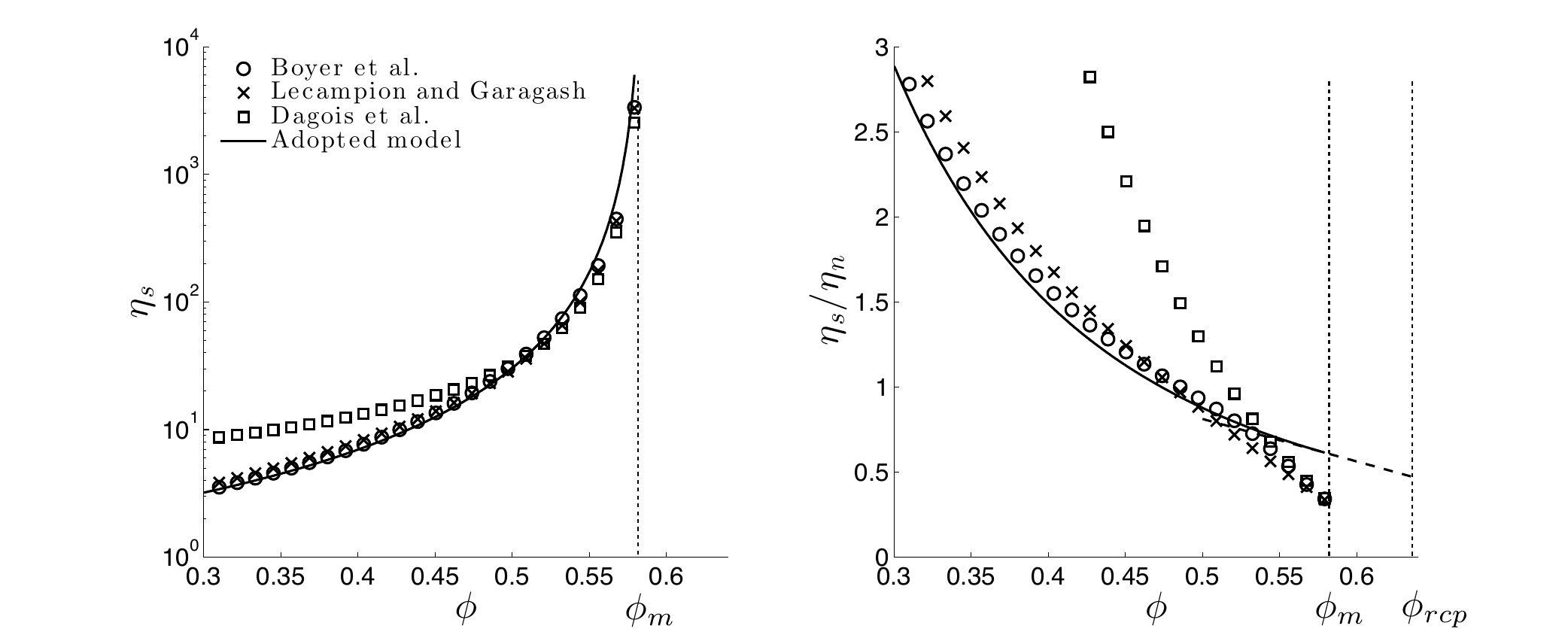}} 
\caption{Effective shear viscosity of suspension (left) and the ratio between the shear and normal viscosities (right) according to the model~\cite{Boyer2011} (circular markers), model~\cite{Lecam2014} (crosses), model~\cite{Dago2015} (square markers), and model~\cite{Morr1999} with $K_s\!=\!0.6$ and $K_n\!=\!1$ (black line).}
 \label{fig1}
\end{figure}

To select a suitable model for further analysis, it is first useful to quantify the differences among them. To fulfill the goal, the left panel in Fig.~\ref{fig1} plots the effective shear viscosity versus particle concentration for a suspension consisting of Newtonian fluid and spherical particles, as defined in~(\ref{taupp}). In particular, predictions of different models are compared. The circular markers show predictions of the model~(\ref{Boymodel})~\cite{Boyer2011}, the crosses show prediction of the model~(\ref{Lmodel}), and the square markers show the prediction of the model~(\ref{Morrmodel3}). The black lines show predictions of the model~(\ref{Morrmodel}) with $K_s\!=\!0.6$ and $K_n\!=\!1$ and is referred to as the ``adopted model''. The value of $K_s\!=\!0.6$ for the model~(\ref{Morrmodel}) was found by minimizing the difference of $\eta_s(\phi)$ with respect to the model~(\ref{Boymodel}). Fig.~\ref{fig1} indicates that all the models predict nearly identical behavior in shear. Note that the model~(\ref{Morrmodel3}) was calibrated against experimental data only for $\phi\gtrsim0.45$, in which case its deviating behavior for small particle concentrations is not surprising. 

The right panel in Fig.~\ref{fig1} plots the ratio between the shear and normal viscosities. One can clearly see that the model~(\ref{Lmodel}) was fitted to~(\ref{Boymodel}). However, the model~(\ref{Morrmodel3}), which is also based on the experimental observations, predicts a different behavior. This is caused by inconsistency of particle pressure measurements in~\cite{Boyer2011} and~\cite{Dago2015}. This inconsistency indicates that the value of $K_n$ is likely to be inaccurate. Luckily, it will become apparent that the value of $K_n$ does not influence the model for suspension flow. It simply changes magnitude of the particle pressure, but keeps the flow profile and fluxes the same. As a result, we select the value $K_n\!=\!1$ for the adopted model, which is consistent with the measurements in~\cite{Boyer2011}. One of the drawbacks of selecting $K_n\!=\!1$ is the fact that the friction coefficient, computed at the maximum concentration, is different from the commonly accepted value of $\eta_s(\phi_m)/\eta_n(\phi_m)\!\approx \!0.3$~\cite{Boyer2011,Dago2015}. This can be incorporated by either using $K_n\!=\!1.8$, which, as discussed above, is not going to change the final result. Or, alternatively, one can use a more complex expression for $\eta_s$ (e.g. as in~\cite{Boyer2011}) to get the correct result. However, since the pressure discrepancy is localized only at the near maximum concentrations, it is expected that its influence also be localized and be relatively small. Therefore, we select a simpler model~(\ref{Morrmodel}) to reduce complexity of computations. This will allow us to focus more on fluid rheology.

\begin{table}
\caption{Summary of rheological models for suspensions.}
\vspace*{3 mm}
\begin{center}
%
\begin{tabular}{l|l|c|l}
Model&Parameters &Compaction & Fluid rheology\\ \hline\hline 
Morris\&Boulay~\cite{Morr1999}&$K_s\!=\!0.1$, $K_n\!=\!0.75$, $\phi_m\!=\!0.68$  & No &Newtonian\\ 
 Boyer et al.~\cite{Boyer2011}&$\mu_1\!=\!0.32$, $\mu_2\!=\!0.7$, $I_0\!=\!0.005$, $\phi_m\!=\!0.585$& No& Newtonian\\ 
 Lecampion\&Garagash~\cite{Lecam2014}& $\mu_1\!=\!0.3$, $\beta\!=\!0.158$, $\phi_m\!=\!0.585$&Yes&Newtonian\\ 
Dagois-Bohy et al.~\cite{Dago2015}&$\alpha_0\!=\!4.6$, $\beta_0\!=\!6$, $\mu_0\!=\!0.3$, $K_n\!=\!0.75$& No&Herschel-Bulkley\\ 
Adopted model&$K_s\!=\!0.6$, $K_n\!=\!1$, $\phi_m\!=\!0.585$, $\phi_{rcp}\!=\!0.64$&Yes&Herschel-Bulkley
\end{tabular}
\vspace*{5 mm}

\begin{tabular}{l|l|l}
Model& Type of validation&Comparison metrics\\ \hline\hline 
Morris\&Boulay~\cite{Morr1999}&Couette flow~\cite{Phil1992}& Particle distribution for $\phi\!\geqslant\!0.45$\\ 
Boyer et al.~\cite{Boyer2011}& Couette flow~\cite{Boyer2011}& Shear and normal stresses for $\phi\!\gtrsim\!0.4$\\ 
Lecampion\&Garagash~\cite{Lecam2014}& Comparison with~\cite{Boyer2011}, pipe flow~\cite{Oh2015}& Particle and velocity profiles for $\phi\!\geqslant\!0.35$\\ 
Dagois-Bohy et al.~\cite{Dago2015}& Couette flow~\cite{Dago2015} &Shear and normal stresses for $\phi\!\gtrsim\!0.45$\\ 
Adopted model&Comparison with~\cite{Eins1905,Boyer2011,Dago2015}&Shear and normal stresses for $\phi>0$\\ 
\end{tabular}

\end{center} 
\label{tab1}
\end{table}

Following the discussion above, we adopt the constitutive model~(\ref{Morrmodel}) with the parameters $K_s\!=\!0.6$ and $K_n\!=\!1$. However, we extend it to Herschel-Bulkley fluids via the approach presented in~\cite{Dago2015}. So that the model can be summarized as follows:
\begin{eqnarray}\label{adoptedmodel}
\tau&=&\dfrac{\eta_s(\phi)}{{\cal F}(\phi) } (\tau_0+k \dot\gamma^n {\cal F}(\phi)^n),\qquad \eta_s =1+2.5\phi \Bigl(1-\dfrac{\phi}{\phi_m}\Bigr)^{-1} +K_s\dfrac{\phi^2}{(\phi_m-\phi)^2},\notag\\ \pp&=&\dfrac{\eta_n(\phi)}{{\cal F}(\phi) } (\tau_0+k \dot\gamma^n{\cal F}(\phi)^n),\qquad \eta_n =K_n\dfrac{\phi^2}{(\phi_m-\phi)^2},\qquad {\cal F}(\phi) =\dfrac{\eta_s(\phi)^{1/2}}{(1-\phi)^{3/2}}.
\end{eqnarray}
The parameters of the model are $K_s\!=\!0.6$, $K_n\!=\!1$, and $\phi_m\!=\!0.585$. Note that the function ${\cal F}(\phi)$ is modified from~(\ref{calF}) to ensure that ${\cal F}(0)=1$, i.e. it applies for small particle concentrations as well. This modification does not significantly alter the behavior at the near maximum concentrations and thus will still match the experimental observations in~\cite{Dago2015}. To model compaction of particles beyond $\phi_m$, we extend the approach~\cite{Lecam2014} within the framework of the adopted model. In particular, equations~(\ref{adoptedmodel}) can be used to define the friction coefficient as $\mu(\phi)\!=\!\tau/\pp\!=\!\eta_s(\phi)/\eta_n(\phi)$. The continuity of the friction $\mu$ and its derivative $d\mu/d\phi$ beyond $\phi_m$ yields to
\begin{equation}\label{compaction}
\tau=\mu_c(\phi)\pp,\qquad \mu_c(\phi)=\dfrac{1}{K_n}\Bigl[K_s+2.5(\phi_m-\phi)\Bigr], \qquad \phi_m\leqslant\phi\leqslant\phi_{rcp},
\end{equation}
which describes compaction until the concentration of random close packing $\phi_{rcp}\!=\!0.64$ is reached. This continuation is shown by the dashed black line in Fig.~\ref{fig1}. Finally, within the compacted region, the shear stress is such that
\begin{equation}\label{nyield}
\tau\leqslant \mu_c(\phi_{rcp}) \pp,\qquad \phi=\phi_{rcp}.
\end{equation}
The modified compaction model~(\ref{compaction}) indicates that the maximum particle concentration is reached at a non-zero value of shear stress, which is in contrast to~(\ref{Lmodel2}) that predicts that $\phi_{rcp}$ corresponds to exactly zero shear stress. Thus, the suggested model~(\ref{adoptedmodel})--(\ref{nyield}) has a potential to provide a better match to the experimental observations in~\cite{Oh2015} since it will model a finite particle plug with concentration $\phi_{rcp}$. 

For completeness, table~\ref{tab1} summarizes features, parameters, and validation types of the described rheological models for suspensions.

\section{Problem formulation for suspension flow in a channel}\label{secchannel}

\subsection{Problem statement}

\begin{figure}[h]
\centerline{\includegraphics[width=0.9\linewidth]{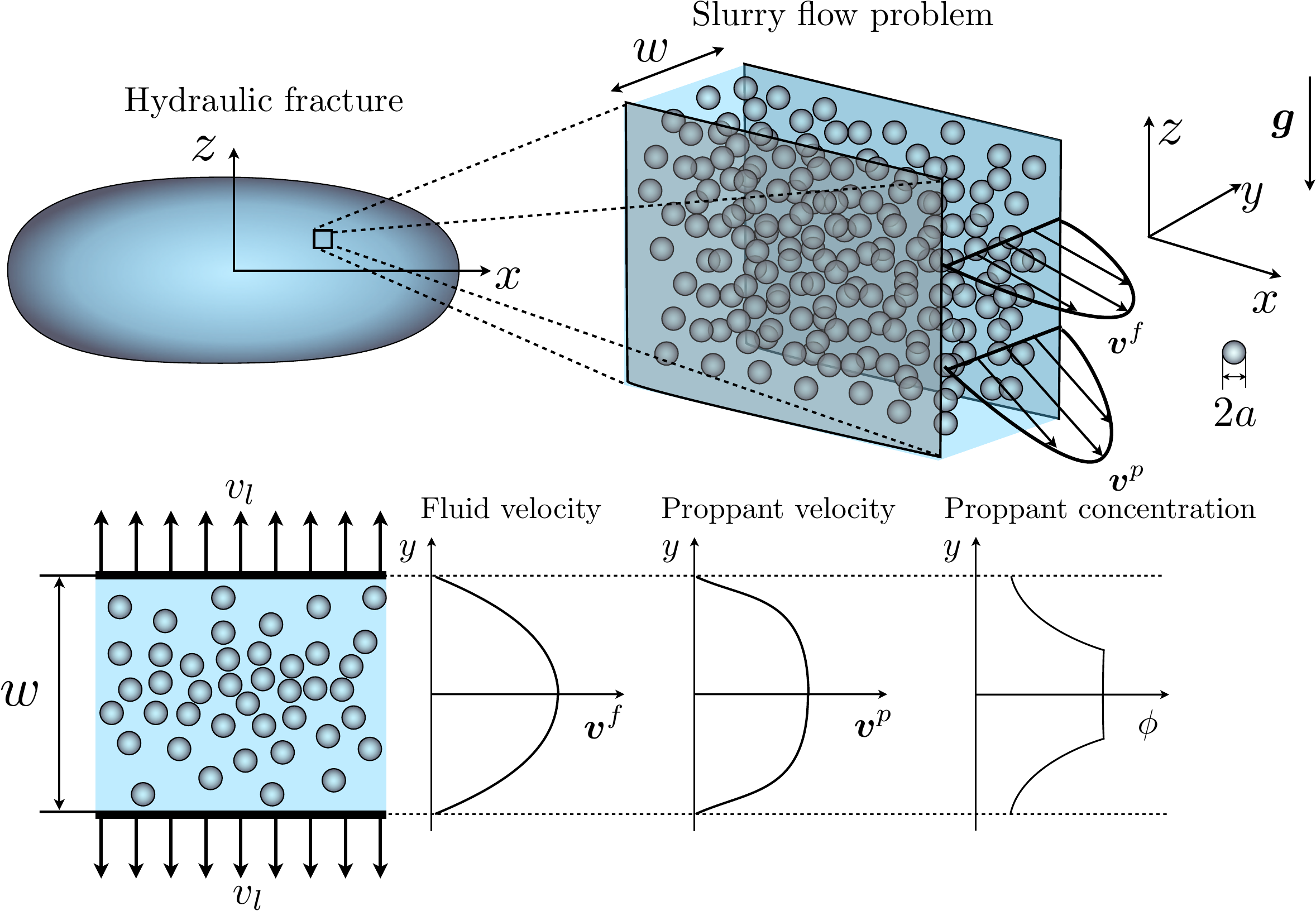}} 
\caption{Relation between hydraulic fracture and suspension flow in a channel. The problem consists of finding the particle velocity profile $\bvp$, the fluid velocity profile $\bvf$, and particle concentration across the channel $\phi$. Fluid is assumed to have Herschel-Bulkley rheology and fluid leak-off with the velocity $v_l$ on the fracture walls is considered.}
 \label{fig2}
\end{figure}

To develop a proppant transport model for hydraulic fracturing, the slurry flow problem in a vertical channel needs to be solved. Fig.~\ref{fig2} shows schematics of a hydraulic fracture and the associated slurry flow problem. Solution for the latter problem is to be used at every point within the fracture to model time evolution of the proppant content during the hydraulic fracture growth. The slurry flow problem consists of finding the particle velocity profile $\bvp$, the fluid velocity profile $\bvf$, and particle concentration across the channel $\phi$ for a given fluid pressure gradient $\nabla p^f$, gravitational acceleration $\boldsymbol{g}$, fracture width $w$, fluid and proppant mass densities $\rhof$ and $\rhop$, particle radius $a$, average proppant concentration $\langle \phi\rangle$, and rheological properties of the carrying fluid. The hydraulic fracture is assumed to be contained in the vertical $(x,z)$ plane, in which case the fluid and particle velocities along the fracture have two components $x$ and $z$, while, at the same time, they vary across the channel, i.e. depend on $y$. We also consider the leak-off velocity $v_l$, which introduces an additional $y$ component of velocity at the fracture surface. Note that only fluid is allowed to leak into the porous formation. The proppant stays within the fracture. The gravitational force acts in the negative $z$ direction and causes proppant settling in that direction. This is one of the reasons why fluid and particle velocities should be considered separately. Finally, proppant volume fraction is also part of the solution and is allowed to vary across the fracture, i.e. $\phi(y)$, as indicated in Fig.~\ref{fig2}.

\subsection{Governing equations}

The balances of linear momentum and mass for the particles can be written as
\begin{eqnarray}\label{govp}
\phi \rhop \Bigl( \pd \bvp t +\bvp \!\cdot\!\nabla\bvp \Bigr)&=&\nabla \!\cdot\! {\boldsymbol \sigma}^p+\phi\rhop \bg+\bff,\notag\\
\pd{\phi\rhop}{t}+\nabla\!\cdot\!(\phi\rhop\bvp)&=&0,
\end{eqnarray} 
where $\phi$ is the volume fraction of the particles, $\rhop$ is the particle mass density, $\bvp$ is the particle velocity, ${\boldsymbol \sigma}^p$ is the particle stress tensor, $\bg$ is the gravity force per unit mass, and $\bff$ is the interaction force between the viscous fluid and the particles. The balances of linear momentum and mass for the fluid phase are
\begin{eqnarray}\label{govf}
(1\!-\!\phi) \rhof \Bigl( \pd \bvf t +\bvf \!\cdot\!\nabla\bvf \Bigr)&=&\nabla\!\cdot\!{\boldsymbol\sigma}^{\textrm{f}}+(1\!-\!\phi)\rhof \bg-\bff,\notag\\
\pd{(1\!-\!\phi)\rhof}{t}+\nabla\!\cdot\!\bigl((1\!-\!\phi)\rhof\bvf\bigr)&=&0,
\end{eqnarray} 
where $\rhof$ and $\bvf$ denote the mass density and fluid velocity respectively, and ${\boldsymbol \sigma}^{\textrm{f}}$ is the stress tensor for fluid. Note that the mass density of the whole mixture is $\rhos=(1\!-\!\phi) \rhof +\phi\rhop$, while the total velocity is $\bvs=(1\!-\!\phi) \bvf +\phi\bvp$. The remaining task is to specify ${\boldsymbol \sigma}^{\textrm{p}}$, ${\boldsymbol \sigma}^{\textrm{f}}$, and $\bff$.

The adopted rheological model~(\ref{adoptedmodel}) provides the expression for the total shear stress of the suspension, but it does not provide a way to specify the contribution of particles and fluid. By employing the concept of local shear rate, which was used to obtain~(\ref{adoptedmodel}), we use the following expressions for the fluid and particle stress tensors
\begin{equation}
{\boldsymbol \sigma}^f=-\pf {\boldsymbol I}+\tau^f \dfrac{2\nabla^s \bvf}{\gammaf},\qquad {\boldsymbol \sigma}^p=-\pp \boldsymbol{Q}+\tau^p \dfrac{2\nabla^s \bvp}{\gammap},
\end{equation}
where $\pf$ is the applied fluid pressure, while the shear stresses are given by
\begin{equation}
\left\{\begin{aligned} \tau^f&=\dfrac{\tau_0+k(\gammaf{\cal F}(\phi))^n}{{\cal F}(\phi)},\qquad \phi<\phi_m,\\
 \tau^f&=0,\qquad \phi\geqslant \phi_m,
\end{aligned}\right.
\qquad 
\left\{\begin{aligned} 
\tau^p &=\dfrac{\eta_s(\phi)\!-\!1}{\eta_n(\phi)} \pp,\qquad \phi<\phi_m,\\
 \tau^p&=\mu_c(\phi) \pp,\qquad \phi_m\leqslant\phi\leqslant\phi_{rcp},\\
 \tau^p&\leqslant\mu_c(\phi_{rcp}) \pp,\qquad \phi=\phi_{rcp}, \end{aligned}\right.\label{sigmap}
\end{equation}
where the shear rates are defined as
\begin{equation}
\gammaf= \sqrt{2\nabla^s \bvf\!:\!\nabla^s\bvf},\qquad  \gammap= \sqrt{2\nabla^s \bvp\!:\!\nabla^s\bvp}.
\end{equation}
Here it is assumed that both fluid and particles are incompressible, $\nabla^s=\tfrac{1}{2}(\nabla +\nabla^T)$ is the symmetric part of the gradient operator. Note that it is also assumed that $|\bvp-\bvf|\ll \bvs$ to have $\nabla\!\cdot\! \bvp=\nabla\!\cdot\!\bvf=0$ from~(\ref{govp}b) and~(\ref{govf}b), which allows us to neglect the divergence terms in calculations of the shear rate. The fluid shear rate is assumed to vanish for concentrations above $\phi_m$, which is the correct limit for $n\!<\!1$. However, $\tau_f$ does not vanish for Newtonian fluids. At the same time, its contribution relative to the particle stress $\tau_p$ is negligible. As follows from~\cite{Morr1999}, the second-order tensor $\boldsymbol{Q}$ describes the anisotropy of the normal stresses and can be represented as
\begin{equation*}
\boldsymbol{Q}~=~\sum _{i=1}^3 \lambda_i \be_i \otimes \be_i,
\end{equation*}
where $\lambda_i=O(1)$~\cite{BoyerPHD} are dimensionless constants and $\be_i$ are unit vectors in the direction of the flow ($i\!=\!1$), gradient ($i\!=\!2$), and vorticity ($i\!=\!3$). Note that $\lambda_2=1$ since the such a component of the stress is measured experimentally and denoted by $\pp$. As follows from~(\ref{adoptedmodel}), the particle pressure is given by
\begin{equation}\label{partpress}
\pp=\dfrac{\eta_n(\phi)}{{\cal F}(\phi) } (\tau_0+k (\gammap{\cal F}(\phi))^n),\qquad \phi<\phi_m.
\end{equation} 
The expressions for $\eta_s(\phi)$, $\eta_n(\phi)$, and ${\cal F}(\phi)$ are specified in~(\ref{adoptedmodel}), while $\mu_c(\phi)$ is given in~(\ref{compaction}).

The interaction force between particles and the fluid $\bff$ has two components: i) the viscous part and ii) buoyancy part, which can be computed from the stress tensor of the fluid. As a result, it can be written as
\begin{equation}\label{bf}
\bff~=~ (1\!-\!\phi)\phi\boldsymbol{f_v}+\phi \nabla\!\cdot\!{\boldsymbol \sigma}^{\textrm{f}},\qquad \boldsymbol{f_{v}}=\dfrac{18 k X(n) }{h_1(\phi) d^{n+1}}\Bigl(h_2(\phi)^{-n} |\bvf\!-\!\bvp|^n+c\dfrac{\tau_0 d^n}{k}\Bigr)\dfrac{\bvf\!-\!\bvp}{|\bvf\!-\!\bvp|},
\end{equation}
where $d\!=\!2a$ is the particle diameter, $X(n)$ is defined in~(\ref{Xn}), $c\!=\!0.823$, and $h_i(\phi)=(1\!-\!\phi)^{\alpha_i}$, $i=1,2$, are the hindrance functions ($\alpha_1\!=\!0.6$ and $\alpha_2\!=\!3.5$). Derivation of the expression for $\boldsymbol{f_v}$ is presented in Appendix A. The expression for $\boldsymbol{f_v}$ represents the body force corresponding to the viscous drag force on a single particle and accounts for interaction between particles through the hindrance functions.  At maximum particle concentrations, on the other hand, it captures the force associated with filtration through packed proppant. 
 
It is important to note that the viscous interaction force $\boldsymbol{f_v}$ is calculated based on the settling velocity of a particle in a quiescent fluid. Equation~(\ref{bf}), on the other hand, will be applied to flowing suspensions. To estimate the effect of flowing suspension, it is necessary to consider typical shear rates associated with the suspension flow and the slip velocity. The shear rate for the suspension can be estimated as $\dot \gamma_{flow} \sim |\bvf|/w$. The shear rate associated with the slip velocity $\dot \gamma_{slip} \sim |\bvf\!-\!\bvp|/a$. The slip velocity is important only if $\bvf\!-\!\bvp=O(\bvf)$ and otherwise can be neglected. In this case $\dot \gamma_{slip}\sim\tfrac{w}{a} \dot \gamma_{flow}$, which upon the assumption $\tfrac{a}{w}\!\ll\! 1$ leads to $\dot \gamma_{slip}\!\gg \!\dot \gamma_{flow}$. As a result, local shear rate associated with the slip velocity dominates and therefore solution for a quiescent fluid is valid. In the situation, in which $\bvf\!-\!\bvp\ll\bvf$, the solution is not valid. However, the error applies to a small quantity (slip velocity) and therefore will not significantly influence the global solution.

\section{Solution for suspension flow in a channel}\label{secchannelsol}

\subsection{Steady slurry flow without slip velocity}

As a starting point, the problem of steady flow of slurry in a channel is considered. To simplify the analysis, the slip velocity between the proppant and fluid is first considered negligible. In addition, the fluid leak-off is ignored at this point as well. The developments are going to proceed in terms of the slurry velocity, defined as
\begin{equation}\label{slurvel}
\bvs=(1\!-\!\phi)\bvf+\phi\bvp.
\end{equation}
Since the slip velocity is neglected, then $\bvp=\bvf=\bvs$. 

In the absence of leak-off velocity, there are only two components of the velocity $v^s_x$ and $v^s_z$, in which case the shear rate for the flow can be calculated as
\begin{equation}\label{srates}
\gammas=\sqrt{\Bigl(\dfrac{\partial v^s_x}{\partial y}\Bigr)^2+\Bigl(\dfrac{\partial v^s_z}{\partial y}\Bigr)^2}.
\end{equation}

By utilizing the assumption of steady flow, the balance of linear momentum for the slurry can be obtained by adding the respective equations for particles and fluid from~(\ref{govp}) and~(\ref{govf}), which leads to
\begin{equation}\label{linmomslur}
0=-\nabla \tilde p^f+\dfrac{\partial }{\partial y} \dfrac{\tau^s}{\gammas} \dfrac{\partial \bvs}{\partial y}+(\phi\!-\!\langle\phi\rangle)(\rhop\!-\!\rhof) \bg,\qquad \tilde p^f=p^f+\rhof g z+\langle \phi\rangle(\rhop\!-\!\rhof)g z,
\end{equation} 
where $\tilde p^f$ is the fluid pressure that has been corrected for the hydrostatic pressure and 
\[
\langle \phi\rangle=\dfrac{2}{w}\int_0^{w/2}\phi\,dy,
\]
is the average proppant concetration. Note that the $z$ coordinate is positive in the upwards direction, which is opposite to the depth direction, i.e. $\boldsymbol {g}=-g\boldsymbol{e}_z$. Hence the sign of the hydrostatic part of the pressure. 

Equations in~(\ref{sigmap}) can be combined to yield 
\begin{eqnarray}\label{shearstressrheol}
\tau^s &=&\dfrac{\eta_s(\phi)}{\eta_n(\phi)} \pp,\qquad \phi<\phi_m,\notag\\
 \tau^s&=&\mu_c(\phi) \pp,\qquad \phi_m\leqslant\phi\leqslant\phi_{rcp},\\
 \tau^s&\leqslant&\mu_c(\phi_{rcp}) \pp,\qquad \phi=\phi_{rcp},\notag
\end{eqnarray} 
where $\pp$ is the particle pressure, which, as per assumption of the steady flow, does not depend on $y$. It is interesting to observe that~(\ref{shearstressrheol}) does not depend on the rheological parameters of the carrying fluid. Relations~(\ref{shearstressrheol}) can be inverted analytically to determine the particle concentrations in terms of $\tau^s/\pp$, i.e.
\begin{equation}\label{Phi}
\phi=\Phi\Bigl(\dfrac{\tau^s}{\pp}\Bigr),
\end{equation}
where the function $\Phi$ has an explicit form.

Equations~(\ref{linmomslur}) and~(\ref{Phi}) should be complemented by the expression for shear stress of the suspension
\begin{equation}\label{taus}
\tau^s=\dfrac{\eta_s(\phi)}{{\cal F}(\phi) } (\tau_0+k (\gammas{\cal F}(\phi))^n),\qquad \gammas>0,
\end{equation} 
which applies only for the flowing part of the suspension.


To proceed with the solution of~(\ref{linmomslur}), (\ref{Phi}), and~(\ref{taus}) it is useful to rewrite these equations in the dimensionless form. In order to do that, let us introduce the following dimensionless parameters
\begin{eqnarray}\label{dimpar}
\Pi^p&=&\dfrac{2\pp}{|\nabla \tilde p^f| w},\qquad T^s=\dfrac{2\tau^s}{|\nabla \tilde p^f| w},\qquad T_0=\dfrac{2\tau_0}{|\nabla \tilde p^f| w},\notag\\
 \dot \Gamma^s&=&\gammas \Bigl(\dfrac{2k}{|\nabla \tilde p^f| w}\Bigr)^{1/n},\qquad \boldsymbol{V}^s=\dfrac{2\bvs}{w}  \Bigl(\dfrac{2k}{|\nabla \tilde p^f| w}\Bigr)^{1/n},\qquad s=\dfrac{2y}{w},\\
G_\rho&=&\dfrac{(\rhop\!-\!\rhof) g}{|\nabla \tilde p^f|},\qquad (\sin\psi,\cos\psi)=-\dfrac{\nabla \tilde p^f}{|\nabla \tilde p^f|},\qquad |\nabla \tilde p^f|=\sqrt{\Bigl(\dfrac{\partial  \tilde p^f}{\partial x}\Bigr)^2+\Bigl(\dfrac{\partial  \tilde p^f}{\partial z}\Bigr)^2}. \notag
\end{eqnarray} 
Here $\psi$ represents the angle between the negative pressure gradient and the vertical $z$ axis. The governing equations can be summarized in the dimensionless form as
\begin{eqnarray}
0&=&\sin\psi+\dfrac{\partial }{\partial s} \dfrac{T^s}{\dot\Gamma^s} \dfrac{\partial V_x^s}{\partial s},\label{goveqdim1}\\
0&=&\cos\psi+\dfrac{\partial }{\partial s} \dfrac{T^s}{\dot\Gamma^s} \dfrac{\partial V_z^s}{\partial s}-(\phi\!-\!\langle\phi\rangle)G_\rho,\label{goveqdim2}\\
T^s&=&\dfrac{\eta_s(\phi)}{{\cal F}(\phi) } (T_0+ (\dot \Gamma^s{\cal F}(\phi))^n),\qquad \dot\Gamma^s>0,\label{goveqdim3}\\
\phi&=&\Phi\Bigl(\dfrac{T^s}{\Pi^p}\Bigr),\label{goveqdim4}\\
\dot \Gamma^s&=&\sqrt{\Bigl(\dfrac{\partial V^s_x}{\partial s}\Bigr)^2+\Bigl(\dfrac{\partial V^s_z}{\partial s}\Bigr)^2}.\label{goveqdim5}
\end{eqnarray} 
We are seeking for a solution of the above system of equations in the form $\Pi^p(\langle\phi\rangle,G_\rho,\psi)$ (particle pressure is constant across the channel), $T^s(s,\langle\phi\rangle,G_\rho,\psi)$, $\phi(s,\langle\phi\rangle,G_\rho,\psi)$,  $V_x^s(s,\langle\phi\rangle,G_\rho,\psi)$, and $V_z^s(s,\langle\phi\rangle,G_\rho,\psi)$.

To initialize the solution of the above governing equations, let us calculate the shear stress from~(\ref{goveqdim1}), (\ref{goveqdim2}), and~(\ref{goveqdim5}) as
\begin{equation}\label{shearstress}
T^s=s \Bigl[1-2G_\rho\cos\psi A(s)+G_\rho^2  A(s)^2\Bigr],\qquad A(s)=\dfrac{1}{s}\int_0^s (\phi\!-\!\langle\phi\rangle)\,ds',
\end{equation} 
where it is used that $T^s(s\!=\!0)\!=\!0$ due to symmetry. This shear stress depends on particle concentration $\phi$ and average particle concentration $\langle\phi\rangle$. Upon substituting equation~(\ref{shearstress}) into the~(\ref{goveqdim4}) we obtain
\begin{equation}\label{phigov}
\phi=\Phi\Bigl(s \Bigl[1-2G_\rho\cos\psi A(s)+G_\rho^2  A(s)^2\Bigr]/\Pi^p\Bigr),\qquad \langle \phi\rangle=\int_0^{1}\phi\,ds.
\end{equation} 
Assuming that the particle pressure is constant, the above system of equations can be solved numerically for $\Pi^p(\langle\phi\rangle,G_\rho,\psi)$ and $\phi(s,\langle\phi\rangle,G_\rho,\psi)$. The shear rate can be calculated from~(\ref{goveqdim3}) as
\begin{equation}\label{Gammas}
\dot \Gamma^s={\cal F}(\phi)^{-1}\Bigl(\max\Bigl\{\dfrac{{\cal F}(\phi) } {\eta_s(\phi)}T^s\!-\!T_0,0\Bigr\}\Bigr)^{1/n},\qquad \phi<\phi_m.
\end{equation} 
The velocity components can be computed from~(\ref{goveqdim1}) and~(\ref{goveqdim2}) as  
\begin{equation}\label{VxVz}
V_x^s=V_{1}^s\sin\psi,\qquad V_z^s=V_1^s\cos\psi-G_\rho V_2^s,
\end{equation}
where the velocity contributions due to pressure gradient and gravity are
\begin{equation}\label{VpVg}
V_1^s=\int_s^1 \dfrac{\dot\Gamma^s}{T^s}s' \,ds',\qquad V_2^s=\int_s^1 \dfrac{\dot\Gamma^s}{T^s} A(s') s' \,ds'.
\end{equation} 
Finally, the flux components can be computed by integrating~(\ref{VxVz}), which gives
\begin{eqnarray}\label{QxQz0}
Q^s_x&=&Q_1^s\sin\psi,\qquad Q^s_z~=~Q_1^s\cos\psi-G_\rho Q_2^s,\\
Q^p_x&=&Q_1^p\sin\psi,\qquad Q^p_z~=~Q_1^p\cos\psi-G_\rho Q_2^p,
\end{eqnarray} 
where the contributions due to pressure gradient and gravity are defined as
\begin{eqnarray}\label{QxQz}
Q_1^s&=&2 \int_0^1 V_1^s \,ds',\qquad Q_2^s~=~2 \int_0^1 V_2^s \,ds',\\
Q_1^p&=&2 \int_0^1 \phi V_1^s \,ds',\qquad Q_2^p~=~2 \int_0^1 \phi V_2^s \,ds'.
\end{eqnarray} 
Here the factor of two accounts for the symmetry of the problem with respect to the center of the channel. Note that the quantities $Q_1^s$, $Q_2^s$, $Q_1^p$, and $Q_2^p$ depend on the average particle concentration $\langle \phi\rangle$, parameter that quantifies the effect of gravity $G_\rho$, and the direction of the pressure gradient that is governed by the angle $\psi$.

It is instructive to consider the limiting case of small particle concentration $\phi\!\ll\! 1$ and $G_\rho\!=\!0$, $\psi\!=\!\pi/2$, i.e. horizontal flow. In this case, ${\cal F}(\phi)\approx1$, $\eta_s(\phi)\approx 1$, and $T^s=s$. As a result, the velocity becomes
\begin{equation}\label{VxVz2}
V_x^s= \int_s^1 \dot\Gamma^s \,ds'=\int_s^1 \Bigl(\max\Bigl\{s\!-\!T_0,0\Bigr\}\Bigr)^{1/n} \,ds'=\dfrac{n}{n\!+\!1}\Bigl(1-T_0\Bigr)^{(n\!+\!1)/n}-\dfrac{n}{n\!+\!1}\Bigl(\max\{s-T_0,0\}\Bigr)^{(n\!+\!1)/n},
\end{equation} 
which corresponds to the velocity profile of Herschel-Bulkley fluid without particles. The flux can be computed as
\begin{equation}\label{QHB}
Q_{HB}=\dfrac{2n}{2n\!+\!1}\Bigl(1-T_0\Bigr)^{(n\!+\!1)/n}\Bigl(1+\dfrac{n}{n\!+\!1}T_0\Bigr).
\end{equation} 
The above expression will be used as a normalization factor to plot the results. 

\subsection{Results}

\begin{figure}[h]
\centerline{\hspace*{-1cm}\includegraphics[width=1.1\linewidth]{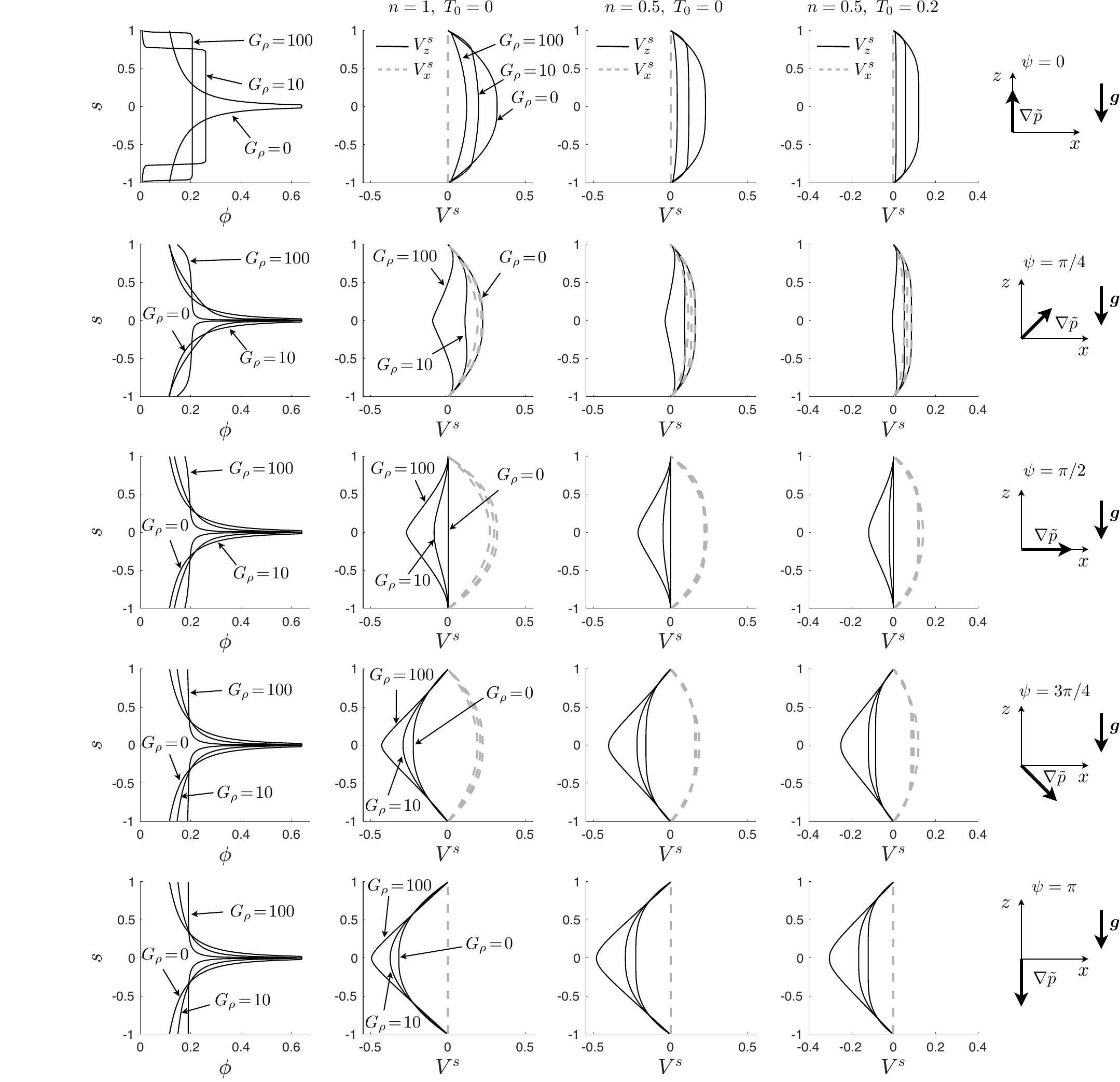}} 
\caption{Spatial variation of the particle concentration and slurry velocity across the channel for different directions of the pressure gradient ($\psi$) and parameter $G_\rho$ for the average particle concentration $\langle \phi \rangle\!=\!0.2$. The first column of plots shows particle concentration, while next three columns contain solutions for velocity profiles for different fluid rheologies: Newtonian, power-law, and Herschcel-Bulkley.}
 \label{figres1}
\end{figure}

Fig.~\ref{figres1} shows numerically computed particle concentration and velocity profiles for different values of problem parameters: $G_\rho$, $n$, $T_0$, and $\psi$. Three fluid rheologies are considered: Newtonian ($n\!=\!1$, $T_0\!=\!0$), power-law ($n\!=\!0.5$, $T_0\!=\!0$), and Herschcel-Bulkley ($n\!=\!0.5$, $T_0\!=\!0.2$). Solution for particle concentration does not depend on fluid rheology and is plotted in the first column. Average particle concentration is taken as $\langle \phi \rangle\!=\!0.2$.

Results in Fig.~\ref{figres1} demonstrate that the equilibrium particle concentration strongly depends on the direction of the pressure gradient for large values of $G_\rho$. This effect is related to the fact that slurry density varies with particle concentration, which in combination with non-uniform particle concentration across the channel leads to a non-uniform slurry density across the channel. The non-uniform density leads to a non-trivial effect of gravity for large values of $G_\rho$, i.e. when the pressure gradient is relatively small compared to the gravitational forces.

 An interesting result is observed for the flow ``against'' gravity for large $G_\rho$ (first row of plots). Particles form a plug with concentration that is below the maximum value. This causes particle pressure to vanish and leads to zero particle concentration away from the plug. This phenomenon is related to the fact that slurry density depends on particle concentration and hence the slurry is heavier in the center. If $G_\rho$ is large, then the pressure gradient is relatively weak and the concentration in the plug is determined by the maximum slurry density that the pressure gradient can ``hold''. Larger densities would lead to local sinking of the slurry. From the first equation in~(\ref{linmomslur}), we obtain
\[
\nabla \tilde p=(\phi_p\!-\!\langle \phi \rangle)(\rho^p\!-\!\rho^f)\boldsymbol{g},
\]
where $\phi_p$ is the particle concentration in the plug. By taking $\psi\!=\!0$ and using the corresponding dimensionless equation~(\ref{goveqdim2}), one has
\begin{equation}\label{phip}
(\phi_p\!-\!\langle \phi \rangle) G_\rho=1,\qquad \phi_p=\langle \phi \rangle+\dfrac{1}{G_\rho}.
\end{equation}
This result agrees with the numerical calculations shown in Fig.~\ref{figres1}. Indeed, particle concentration in the plug is $\phi_p\!=\!0.3$ for $G_\rho\!=\!10$ (recall that $\langle\phi\rangle\!=\!0.2$). Equation~(\ref{phip}) also allows us to predict the critical value of $G^*_\rho$, at which this phenomenon occurs
\[
G_\rho^*=\dfrac{1}{\phi_{rcp}\!-\!\langle \phi \rangle}.
\]
The corresponding shear rate is zero within the plug region, which leads to a blunted velocity profile. This behavior is present in all fluid types considered.

Variation of the slurry density across the channel also influences the flow in other directions. Even for the horizontal pressure gradient, such density variation leads to sinking of the central part of the flow (see the third row of plots in Fig.~\ref{figres1}). This introduces a downward particle transport mechanism even in the absence of slip velocity between the phases. For the pressure flow ``along the gravity'', higher density in the center of the channel leads to a sharper velocity profile for large values of $G_\rho$. The effect is more pronounced for power-law and Herschel-Bulkley fluids since the gravity increases shear stress and shear rate near the center, which effectively reduces apparent values of local viscosity of the slurry.

\begin{figure}[h]
\centerline{\hspace*{-1cm}\includegraphics[width=0.9\linewidth]{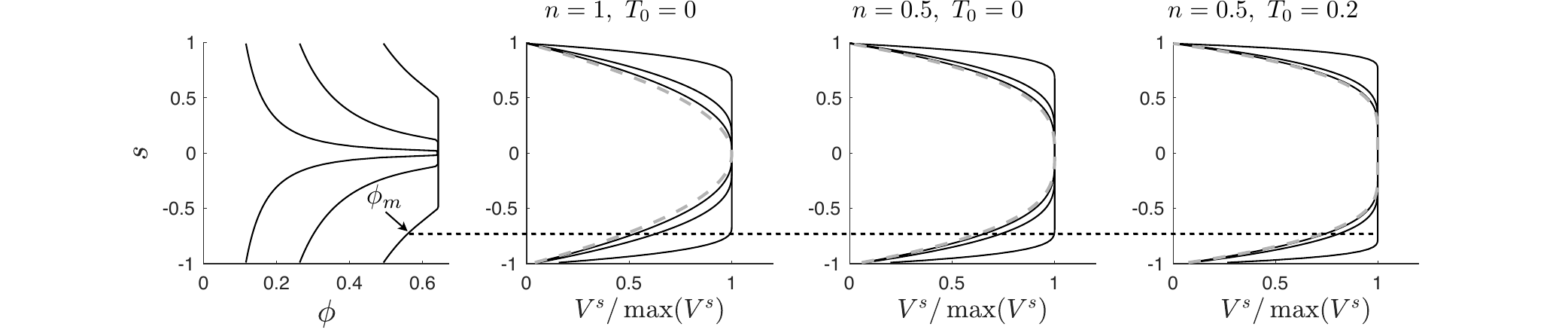}} 
\caption{Spatial variation of particle concentration and slurry velocity across the channel for buoyant particles and particle concentration $\langle \phi \rangle\!=\!\{0.2, 0.4,0.6\}$. The first  plot shows particle concentration, while next three plots display solutions for velocity profiles for different fluid rheologies: Newtonian, power-law, and Herschcel-Bulkley. Dashed gray lines indicate velocity solution without particles.}
 \label{figresbuoyant}
\end{figure}

Fig.~\ref{figresbuoyant} shows particle concentration and velocity solutions for buoyant particles, i.e. for $G_\rho\!=\!0$ and different particle concentrations $\langle \phi \rangle\!=\!\{0.2, 0.4,0.6\}$. Velocity profiles corresponding to the case without particles are shown by the dashed gray lines. According to the model, particles form a plug with concentration $\phi_{rcp}$ in the center of the channel. The shear stress in this zone is not sufficient to initiate dilation (the third equation in~(\ref{shearstressrheol})). Then, particle concentration gradually decreases to $\phi_m$, which corresponds to the compaction zone, which is governed by the second equation in~(\ref{shearstressrheol}). The shear rate is still zero in this zone, so that the velocity profile is flat. After that, concentration decreases gradually and suspension starts to yield. Note that if the fluid yield stress is present, then the slurry starts to experience shear motion for even smaller particle concentrations once the yield stress of fluid is exceeded.

\begin{figure}[h]
\centerline{\hspace*{-1cm}\includegraphics[width=0.9\linewidth]{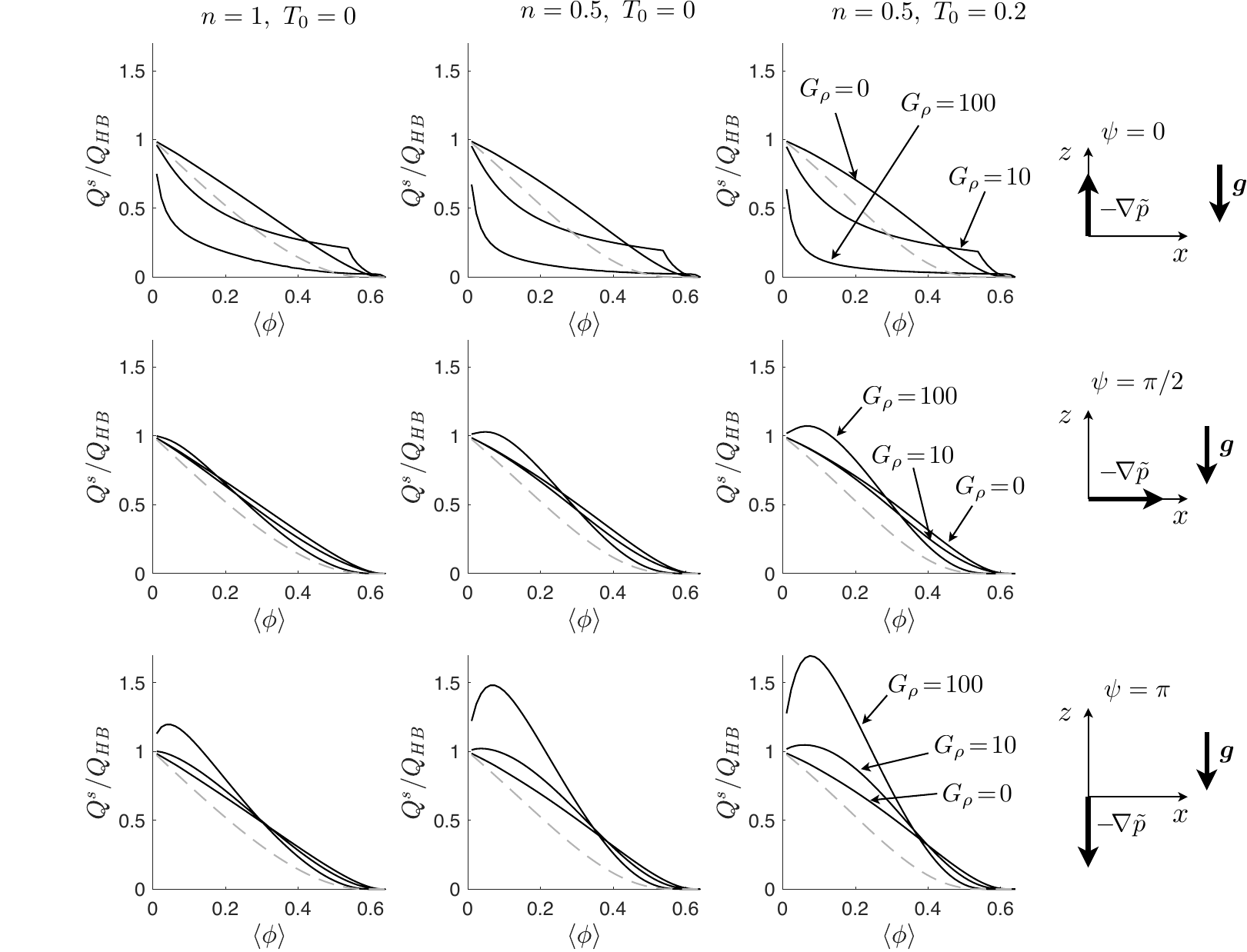}} 
\caption{Variation of the slurry flux normalized by the flux of clean fluid versus average particle concentration for different directions of the pressure gradient ($\psi$) and parameter $G_\rho$. The columns contain solutions for different fluid rheologies: Newtonian, power-law, and Herschcel-Bulkley.}
 \label{figQs}
\end{figure}

To investigate the effect of gravity on the slurry flux, Fig.~\ref{figQs} shows the variation of the total slurry flux, normalized by the solution for clean fluid~(\ref{QHB}), versus average particle concentration. Here the total flux is defined as $Q_s=\sqrt{(Q^s_x)^2\!+\!(Q^s_z)^2}$, where the components are calculated via~(\ref{QxQz0}). Three fluid rheologies are considered and the effect of $G_\rho$ and direction of the pressure gradient are investigated. Dashed gray lines in Fig.~\ref{figQs} indicate solution that corresponds to uniform particle concentration. The ratio between the fluxes that is plotted in Fig.~\ref{figQs} has the meaning of inverse apparent viscosity. The suspension becomes effectively thicker if the ratio is less then one and thinner if the ratio is greater than one. The slurry flux follows a gradual trend for the horizontal direction of the pressure gradient. More interesting results are observed for the vertical directions. For the upward direction, there is a characteristic value of average concentration, above and below which there are qualitatively different solutions. This corresponds to the transition to the solution shown in the first row of plots in Fig.~\ref{figres1}, for which particle concentration is uniform in the central part and is zero outside of it. The value for the average concentration that corresponds to the transition point can be calculated from~(\ref{phip}) by taking $\phi_p\!=\!\phi_{rcp}$, i.e. $\langle\phi\rangle^*\!=\!\phi_{rcp}-1/G_\rho$. For the downward direction, large values of $G_\rho$ lead to ``thinning'' of the suspension for small particle concentrations. This can be traced back to the velocity profiles shown in Fig.~\ref{figres1} (last row), which demonstrate that $G_\rho$ leads to elongated velocity profiles with larger maximum velocity. 

\begin{figure}[h]
\centerline{\hspace*{-1cm}\includegraphics[width=0.9\linewidth]{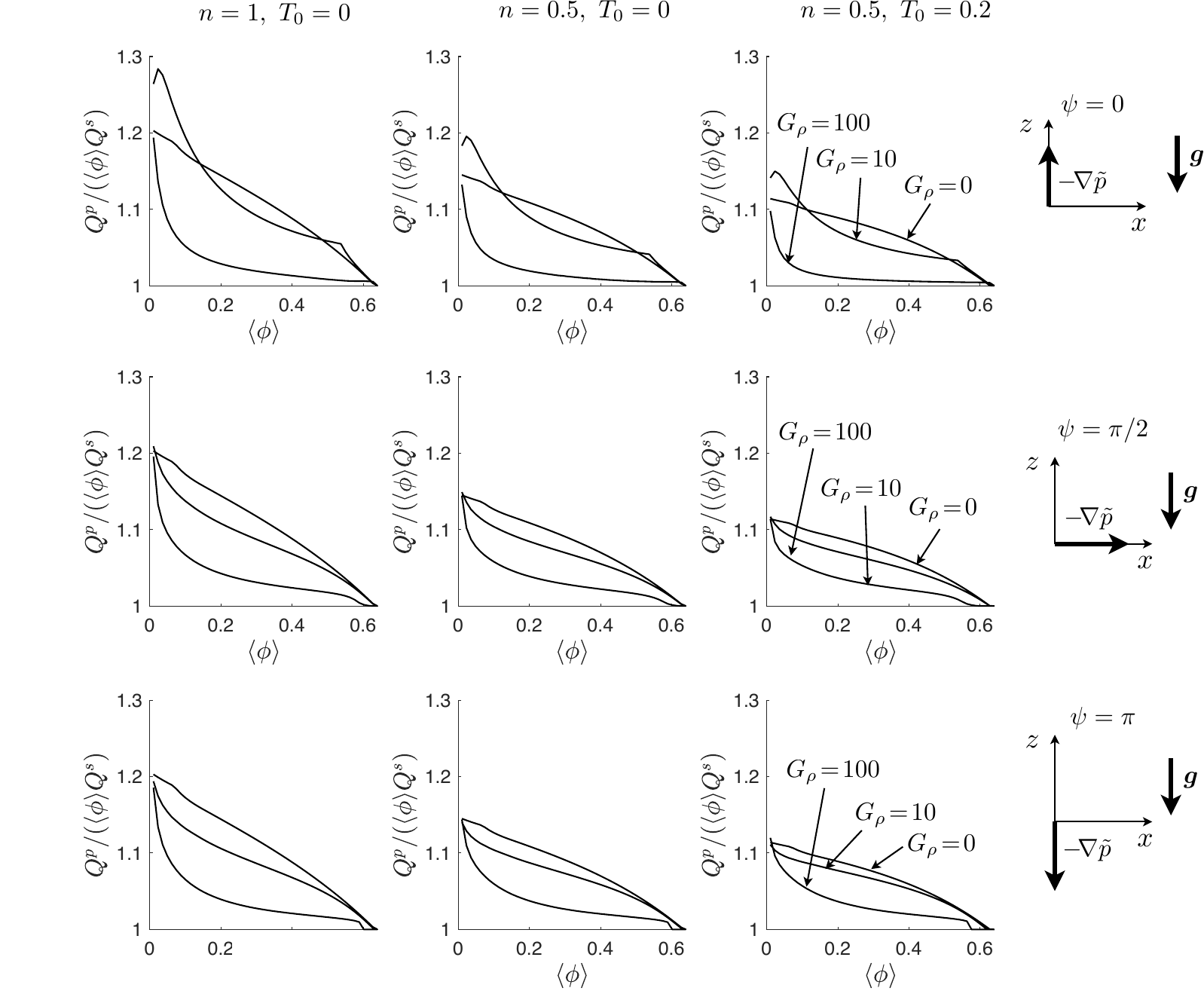}} 
\caption{Variation of the proppant flux normalized by the slurry flux multiplied by average particle concentration (which is proppant flux corresponding to uniform particle variation across the channel) versus average particle concentration for different directions of the pressure gradient ($\psi$) and parameter $G_\rho$. The columns contain solutions for different fluid rheologies: Newtonian, power-law, and Herschcel-Bulkley.}
 \label{figQp}
\end{figure}

To quantify average particle transport, Fig.~\ref{figQp} displays variation of the quantity $Q^p/(\langle\phi\rangle Q^s)$ versus average particle concentration. Here $Q_p=\sqrt{(Q^p_x)^2\!+\!(Q^s_p)^2}$ is the total flux of particles, whose components are defined in~(\ref{QxQz0}). The normalization factor $\langle\phi\rangle Q^s$ represents particle flux in the situation, in which particles are distributed uniformly across the channel. Therefore, the quantity $Q^p/(\langle\phi\rangle Q^s)$ represents the change of the particle flux caused by migration of the particles. In addition, it can be interpreted as the ratio between average particle and slurry velocities. Results in Fig.~\ref{figQp} demonstrate that accumulation of particles near the center of the flow leads to faster transport of particles. The quantitative results depend on fluid rhelology, direction of the pressure gradient, and $G_\rho$.

It is very important to emphasize that the obtained solutions for particle concentration (and the associated velocity profiles and fluxes) represent ``equilibrium'' solutions, i.e. solutions that correspond to uniform particle pressure across the channel. The equilibrium may not always be reached in reality. This especially applies to fluids with non-Newtonian rheology. Time scale for particle migration clearly depends on apparent viscosity of the carrying fluid. Non-Newtonian fluids exhibit higher apparent viscosity near the center of the flow and hence particle migration speed will be reduced. For fluids with yield stress particle migration may completely stop once particles reach the zone with shear stress below the yield stress. This may lead to existence of multiple solutions for particle distribution and make the problem history-dependent.

\subsection{Effect of fluid leak-off}

Fluid leak-off at the fracture walls can potentially influence particle concentration profile and therefore can affect the apparent slurry viscosity. Since we are considering steady flow, the particles should have only one component of the velocity along the flow, while fluid has an additional component in the direction towards the fracture walls, which causes a slip velocity $\Delta v_y$. Presence of the slip velocity between particles and fluid induces particle pressure gradient, which can be computed as
\[
\dfrac{\partial \pp}{\partial y}=\phi(1\!-\!\phi)f_{v,y}(\Delta v_y),
\]
where $f_{v,y}$ is the $y$ component of the viscous drag force~(\ref{bf}). On the other hand, if one considers filtration of the fluid through proppant pack, we have $f_{v,x}(v_d) = \tfrac{1}{\phi} \nabla p^f$~(\ref{govsettsol}), where $v_d$ is the filtration velocity. Since $\Delta v_y = O(v_l)$, where $v_l$ is the leak-off velocity and $f_{v,y}(\Delta v_y)/f_{v,x}(v_d)=O(\Delta v_y/v_d)$ (for Newtonian fluids), the pressure drop over the fracture cross-section can be estimated from the above equation as
\[
\Delta \pp=O\Bigl(\dfrac{v_l}{v_d} w\nabla p^f \Bigr),
\]
where the fracture width $w$ is taken as the characteristic length scale in the direction across the channel. It is also important to note that the $\pp=O(\tau^s)$ from~(\ref{sigmap}). The total shear stress, on the other hand, is $\tau^s=O(w\nabla p^f)$. This allows us to write
\[
\Delta \pp=O\Bigl(\dfrac{v_l}{v_d} \Bigr)\pp.
\]
Finally, since permeability of rock is typically much smaller than the permeability of proppant in practical applications (that is the main reason to do hydraulic fracturing), the ratio $v_l/v_d\!\ll \!1$. As a result, the influence of leak-off on particle pressure (and therefore concentration) can be neglected. Note that despite the above estimate considers Newtonian fluids, it can be easily generalized to Herschel-Bulkley fluid. Also, here we consider leak-off to rock matrix. Situations with concentrated leak-off, e.g. associated with permeable interfaces between rock layers, may be different. Due to the assumption of steady flow, however, they cannot be handled in the current framework.


\section{Suspension and particle fluxes with slip velocity}\label{secfluxes}

The previous section considered motion of suspension as a whole and assumed that the slip velocity between the phases is small. Indeed, the slip velocity is small in most cases. However, it becomes important once proppant approaches its packing limit, or when particle settling is significant. Both effects are desirable to account for in practical applications. One possibility is to derive expressions for the slip velocity in a mathematically consistent way from the governing equations~(\ref{govp}) and~(\ref{govf}), which was done in~\cite{Dont2014} for the case of Newtonian fluid. This, however, might be challenging to do precisely for a given non-linear fluid rheology. Alternatively, one may learn from the results of~\cite{Dont2014} and to include the effect of slip velocity approximately, in which we construct the solution that satisfies some known limiting behavior. We adopt the second approach. In particular, the slurry and particle fluxes are taken in the following form 
\begin{eqnarray}
q^s&=&q^s_{noslip}+\dfrac{\langle\phi\rangle}{\phi_{rcp}}q^s_{filt},\label{slurryflux} \\
q^p&=&q^p_{noslip}+q^p_{settl},\label{proppantflux}
\end{eqnarray} 
 where the contribution of the solution without slip velocity can be summarized from previous section as
 \begin{eqnarray}\label{qnoslip}
q^s_{noslip}&=&H_{flow}   \Bigl [-Q^s_1\, \nabla \tilde p^f+ Q^s_2\, (\rhop\!-\!\rhof) \bg\Bigr],\qquad H_{flow}=\dfrac{w^2}{4|\nabla \tilde p^f|} \Bigl(\dfrac{|\nabla \tilde p^f| w}{2k}\Bigr)^{1/n},\label{slurryfluxnoslip} \notag\\
q^p_{noslip}&=&H_{flow} \Bigl [-Q^p_1\, \nabla \tilde p^f+ Q^p_2\, (\rhop\!-\!\rhof) \bg\Bigr],\label{proppantfluxnoslip}
\end{eqnarray} 
 where ${ Q^s_i}(\langle\phi\rangle, G_\rho,\psi,n,T_0)$ and ${ Q_i^p}(\langle\phi\rangle, G_\rho,\psi,n,T_0)$ ($i=1,2$) are the dimensionless functions defined in~(\ref{QxQz0}), while $H_{flow}$ represents hydraulic conductivity associated with the viscous flow between two parallel plates. As can be seen from~(\ref{slurryflux}), the slip velocity introduces the term $q^s_{filt}$ that describes fluid filtration through packed proppant. The contribution of this term is negligible for particle concentrations below packing limit. However, since $q^s_{noslip}$ vanishes at the maximum concentration, the filtration term becomes the only term that contributes to the slurry flux. Therefore, it is important to account for $q^s_{filt}$ only for concentrations that are close to $\phi_{rcp}$, which is represented via the multiplier ${\langle\phi\rangle}/{\phi_{rcp}}$. It ensures that this filtration term vanishes if there are no particles, and fully captures the effect of filtration for the maximum concentrations. The term $q^p_{settl}$ in~(\ref{proppantflux}) describes particle settling due to a mismatch of particle and fluid densities. Expressions for $q^s_{filt}$ and $q^p_{settl}$ are provided below.
 
To compute the filtration flux, we need to combine the viscous force~(\ref{bf}) with the corresponding governing equation obtained in the limit of filtration~(\ref{govfiltsol}) (for which $q^s_{noslip}\!=\!0$). Since the proppant velocity is zero during filtration and the velocity profile is uniform across the fracture, then the filtration slurry flux can be computed as
\begin{equation}\label{qsfilt}
q^s_{filt}= H_{filt} \,(\!-\! \nabla p^f\!+\!\rhof\bg),\qquad H_{filt} = \dfrac{w\, (1\!-\!\phi_{rcp})h_2(\phi_{rcp})}{{|\!-\! \nabla p^f\!+\!\rhof\bg|}} \Bigl[\dfrac{h_1(\phi_{rcp}) d^{n+1}} {18 k X(n)\phi_{rcp} }\, |\!-\! \nabla p^f\!+\!\rhof\bg|-c\dfrac{\tau_0 d^n}{k}\Bigr]^{1/n},
\end{equation} 
where $H_{filt}$ is the hydraulic conductivity that is associated with filtration and we used the fact that the ratio between the slurry flux and fluid flux is $(1\!-\!\phi_{rcp})$. In the limiting case of Newtonian fluids, this expression coincides with the previously obtained result in~\cite{Dont2014}. Note that the pressure gradient in~(\ref{qsfilt}) should be sufficiently large so that the term in square brackets is positive. Otherwise, the hydraulic conductivity is zero.

In order to obtain the settling flux, it is first convenient to rewrite the slurry and particle fluxes as
 \begin{eqnarray}
  q^s&=& -H_{flow}\hat Q_1^s \, \nabla \tilde p^f+ H_{flow} \hat Q^s_2 (\rhop\!-\!\rhof)\bg,\label{slurry2}\\
 q^p&=&\hat Q^p_1\, q^s + H_{flow}\hat Q_2^p\, (\rhop\!-\!\rhof) \bg+q^p_{settl},\label{particle2}
 \end{eqnarray} 
where the newly defined functions are given by
 \begin{eqnarray}
 \hat Q_1^s&=&Q^s_1 + \dfrac{\langle\phi\rangle}{\phi_{rcp}}\,\dfrac{H_{filt}}{H_{flow}},\qquad \hat Q^s_2=  Q^s_2\ -\dfrac{\langle\phi\rangle^2}{\phi_{rcp}}\,\dfrac{H_{filt}}{H_{flow}},\\  
 \hat Q_1^p&=&\dfrac{Q^p_1}{ \hat Q_1^s}, \qquad \hat Q_2^p=Q^p_2 - \hat Q^p_1\hat Q^s_2.
 \end{eqnarray} 
Now, consider the problem of particle settling in a still suspension with uniform particle concentration (i.e. $\phi\!=\!\langle\phi\rangle$) and away from the filtration limit $H_{filt}\!\ll\! H_{flow}$. In this case, we have $\nabla \tilde p^f\!=\!0$ and the total particle settling flux reduces from~(\ref{particle2}) to 
 \begin{equation}\label{totsetflux}
q^p|_{q^s=0}=H_{filt} \dfrac{\langle\phi\rangle^2}{\phi_{rcp}}\dfrac{Q_1^p}{Q_1^s}\, (\rhop\!-\!\rhof) \bg+q^p_{settl}.
 \end{equation} 
 where we used the fact that $Q^p_2 \!=\!Q^s_2\!=\!0$ for uniform particle concentration and $H_{filt}\!\ll \!H_{flow}$ to replace $\hat Q_1^s$ with $Q_1^s$. On the other hand, one can combine the expression for the viscous force~(\ref{bf}) with the corresponding governing equation obtained in the limit of settling~(\ref{govsettsol}), to get
\begin{equation}\label{vsettl}
\bvp\!-\!\bvf = \dfrac{\bvp-\bvs}{1-\langle\phi\rangle}=h_2(\langle\phi\rangle) \Bigl[\dfrac{h_1(\langle\phi\rangle) d^{n+1}}{18 k X(n) }(\rhop\!-\!\rhof) g - c\dfrac{\tau_0 d^n}{k}\Bigr]^{1/n}\,\dfrac{\bg}{g},
\end{equation} 
where $\bvs \!=\! \langle\phi\rangle \bvp \!+\!(1\!-\!\langle\phi\rangle)\bvf$ is the slurry velocity. In order to compute the flux~(\ref{totsetflux}), we need to integrate~(\ref{vsettl}) over the channel width and take $q^s\!=\!\int \bvs\, dy \!=\!0$. As a result, equations~(\ref{totsetflux}) and~(\ref{vsettl}) can be combined to yield
 \begin{equation}\label{qsettl}
q^p_{settl}=H_{settl} G^p\, (\rhop\!-\!\rhof) \bg,
\end{equation} 
where 
 \begin{equation}\label{Gp}
 G^p = 1-\dfrac{\langle\phi\rangle^2}{\phi_{rcp}}\dfrac{H_{filt}}{H_{settl}}\dfrac{Q_1^p}{Q_1^s},\qquad H_{settl}=\dfrac{w\langle\phi\rangle^2(1\!-\!\langle\phi\rangle)h_2(\langle\phi\rangle)}{|\!-\! \nabla p^f\!+\!\rhof\bg|} \Bigl[\dfrac{h_1(\langle\phi\rangle) d^{n+1}}{18 k X(n) \langle\phi\rangle}|\!-\! \nabla p^f\!+\!\rhof\bg|- c\dfrac{\tau_0 d^n}{k}\Bigr]^{1/n}.
\end{equation} 
Here $H_{settl}$ is the hydraulic conductivity that is associated with settling of particles and we used the fact that $\nabla \tilde p^f\!=\!0$ for the considered settling problem to replace $\langle\phi\rangle(\rhop\!-\!\rhof) g$ with $|\!-\! \nabla p^f\!+\!\rhof\bg|$ in~(\ref{Gp}) via~(\ref{linmomslur}). It is important to note that the constructed function $G^p$ always vanishes for the maximum particle concentration, which guarantees that there is no proppant flux (neither due to convection nor due to settling) once the particles have bridged in the channel. Also, it closely matches the result obtained in~\cite{Dont2014b} for Newtonian fluids and using a more rigorous derivation procedure that accounts for non-uniform particle distribution across the channel.

In summary, to compute slurry and particle fluxes, one should first calculate three hydraulic conductivities $H_{flow}$, $H_{filt}$, and $H_{settl}$ using~(\ref{qnoslip}), (\ref{qsfilt}), and~(\ref{Gp}) for given problem parameters. Then, the corresponding fluxes can be evaluated as
 \begin{eqnarray}
  q^s&=& -H_{flow}\hat Q_1^s \, \nabla \tilde p^f+ H_{flow} \hat Q^s_2 (\rhop\!-\!\rhof)\bg,\label{slurry3}\\
 q^p&=&\hat Q^p_1\, q^s + H_{settl}\hat G^p\, (\rhop\!-\!\rhof) \bg,\label{particle3}
 \end{eqnarray} 
where the dimensionless functions are defined as
 \begin{eqnarray}\label{QGfcns}
 \hat Q_1^s&=&Q^s_1 + \dfrac{\langle\phi\rangle}{\phi_{rcp}}\,\dfrac{H_{filt}}{H_{flow}},\qquad \hat Q^s_2=  Q^s_2\ -\dfrac{\langle\phi\rangle^2}{\phi_{rcp}}\,\dfrac{H_{filt}}{H_{flow}},\notag\\  
 \hat Q_1^p&=&\dfrac{Q^p_1}{ \hat Q_1^s}, \qquad \hat Q_2^p=Q^p_2 - \hat Q^p_1\hat Q^s_2,\qquad \hat G^p=1-\dfrac{\langle\phi\rangle^2}{\phi_{rcp}}\dfrac{H_{filt}}{H_{settl}}\dfrac{Q_1^p}{Q_1^s}+\dfrac{H_{flow}}{H_{settl} }\hat Q_2^p,
 \end{eqnarray} 
in which $Q^s_1$, $Q^s_2$, $Q^p_1$, and $Q^p_2$ are computed from the problem of suspension flow without slip velocity~(\ref{QxQz}). To facilitate implementation of the slurry and particle fluxes~(\ref{slurry3}) and~(\ref{particle3}) into a proppant transport module of a hydraulic fracturing simulator, approximate expressions for the functions $Q^s_1$, $Q^s_2$, $Q^p_1$, and $Q^p_2$ are presented in Appendix~\ref{QsQpappr}.

\section{Discussion}\label{secdisc}

\begin{figure}[h]
\centerline{\includegraphics[width=1.1\linewidth]{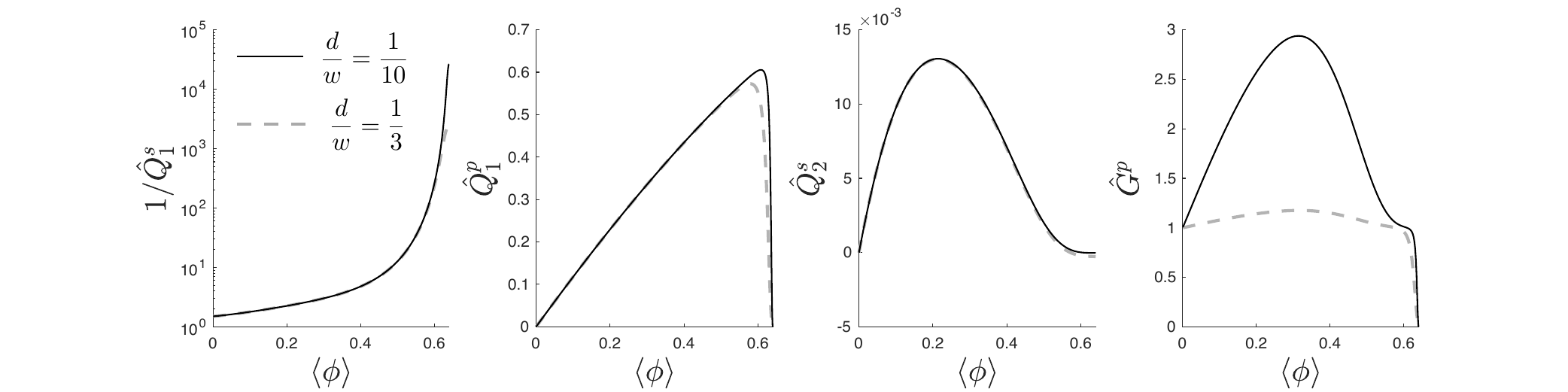}} 
\caption{Variation of the functions $\hat Q_1^s$, $ \hat Q_1^p$, $ \hat Q_2^s$, and $ \hat G^p$ versus average particle volume fraction for two different relative particle sizes $d/w$.}
 \label{QGplot}
\end{figure}

Proppant transport equations presented in this study are derived based on the frictional rheology of suspension carried by Hershcel-Bulkley fluid. They represent a general case and appear to be complicated. To better understand the behavior that is incorporated into the developed model, this section aims to discuss meaning of each of the terms in~(\ref{slurry3}) and~(\ref{particle3}). To aid the description, Fig.~\ref{QGplot} plots the functions $\hat Q_1^s$, $ \hat Q_1^p$, $ \hat Q_2^s$, and $ \hat G^p$ versus average particle volume fraction $\langle \phi\rangle$ for the case of Newtonian fluid and two different relative particle sizes $d/w$, where $d$ is particle diameter and $w$ is the channel width. Two cases are considered $d/w=1/3$ which corresponds to flow prior to bridging and $d/w=1/10$ which corresponds to the flow in the middle of the fracture.

The first panel in Fig.~\ref{QGplot} plots inverse of the function $\hat Q_1^s$, which has the meaning of apparent viscosity, see the first term in~(\ref{slurry3}). One can observe that the apparent viscosity is effectively truncated at very large values due to the presence of filtration in the model that still allows for fluid flow even for the maximum particle volume fraction.

The second plot in Fig.~\ref{QGplot} plots variation of the function  $\hat Q_1^p$. This function reflects proppant advection with the slurry flow, see the first term in~(\ref{particle3}). Behavior for small particle volume fractions is linear and then it abruptly goes to zero. This ensures that the particle flux vanishes due to proppant bridging at maximum concentrations while the slurry flux is non-zero due to filtration. 

The third plot in Fig.~\ref{QGplot} represents the second term in~(\ref{slurry3}) or $ \hat Q_2^s$. This term can be understood as a correction to the hydrostatic pressure gradient due to interaction between the particles and the wall as well as due to non-uniform particle distribution across the channel. As can be seen from the display, it is relatively small and can therefore be neglected for practical computations.

The fourth and the final plot in Fig.~\ref{QGplot} represents the second term in~(\ref{particle3}) or $ \hat G^p$, which is related to particle settling due to density mismatch between the fluid and the particles. There are thee contributions to this term: i) settling due to slip velocity between the phases, ii) correction due to interaction with the wall, and iii) correction due to non-uniform particle distribution across channel, which induces slumping of the center part of the flow caused by density variation across the channel. The plotted quantity $ \hat G^p$ represents the rate of particle settling normalized by the settling rate due to slip velocity only. Therefore, variations from unity correspond to the corrections by the effects of the wall and particle distribution. Interaction with the wall leads to reducing the flux to zero for maximum particle volume fraction. This is caused by the shear force between the particles and the wall that holds the proppant against settling. Correction due to non-uniform particle distribution across the channel seems to be more interesting. Results demonstrate that the settling rate can be effectively increased due to particle accumulation near the center of the channel. The effect is more pronounced for relatively small particles, for which the inherent settling rate due to slip is small. This can also be concluded from~(\ref{particle3}), (\ref{QGfcns}) and the definitions of hydraulic conductivities for the flow and for settling. Component of $\hat G^p$ that is proportional to $\hat Q_2^p$ leads to the particle settling rate that is independent of the particle size. While this term can be small for relatively large particles, it becomes dominant for small particle sizes since other terms scale with the particle size. It should be noted, however, that these results should always be understood in the context of model assumptions. One of which is steadiness of the solution. It is well known that it takes longer for small particles to reach steady solution compared to large particles. As a result, domination of slumping may be delayed in some situations until the steady solution is reached.

\section{Summary}\label{secsummary}

This study presents an investigation of flow of suspensions that consist of Herschel-Bulkley fluid mixed with spherical particles. Review of the existing suspension flow models is presented at the beginning and a suitable model has been selected for further analysis. The governing equations are formulated for the general case of suspension flow and then steady flow in a vertical channel is analyzed in detail. In particular, two-dimensional steady flow is considered, in which both the pressure gradient and the gravitational force drive the suspension flow. Solution for particle concentration variation across the channel as well the velocity profile are computed numerically. It is shown that the non-linear fluid rheology and gravitational force lead to different solutions depending on orientation of the pressure gradient relative to the vertical axis. This is because the non-uniform particle distribution across the width of the channel introduces variations of density of the suspension, which in turn causes sinking of the denser regions. This leads to a mechanism of downward particle motion even without slip velocity between the phases. To aid the implementation of the developed model into a proppant transport module of a hydraulic fracturing simulator, expressions for the slurry and particle fluxes are obtained, which also account for the slip velocity between the phases. In particular, the slip velocity leads to capturing the fluid filtration through the packed proppant pack at the maximum particle concentration and particle settling due to mismatch between particle and fluid densities. In addition, approximate expressions for the slurry and particle fluxes are obtained, which can be used for practical applications.

\appendix

\section{Viscous drag force between particles and Herschel-Bulkley fluid}\label{append1}

\subsection*{Settling velocity of a single rigid sphere in Herschel-Bulkley fluid}

Consider the problem of a single spherical particle with diameter $d$ settling in a Herschel-Bulkley fluid, that is characterized by yield stress $\tau_0$, consistency index $k$, and fluid behavior index $n$. Terminal particle settling velocity $v_z^p$ is determined by the balance between the gravitational and buoyancy forces, and the viscous drag force
\begin{equation}\label{CD}
C_D \dfrac{\rho_f (v_z^p)^2}{2} \dfrac{\pi d^2}{4} =(\rho^p\!-\!\rho^f) g \dfrac{\pi d^3}{6},
\end{equation}
where $\rho^p$ is the density of particle, $\rho^f$ is the density of fluid, $d$ is the particle diameter, $C_D$ is the drag coefficient, and $g$ is the gravitational acceleration. The drag coefficient for the problem under consideration depends on two dimensionless parameters 
\begin{equation}\label{ReBm}
Re=\dfrac{\rho^f (v_z^p)^{2-n}d^n}{k},\qquad Bm=\dfrac{\tau_0 d^n}{k (v_z^p)^n},
\end{equation}
where $Re$ is Reynolds number and $Bm$ is Bingham number. As shown in~\cite{Ansley1967}, for small Reynolds numbers the drag coefficient can be expressed in terms of the so-called dynamic parameter $Q$ as
\begin{equation}\label{CD2}
C_D=\dfrac{24 X(n)}{Q},\qquad Q=\dfrac{Re}{1+c \,Bm}, 
\end{equation}
The numerical coefficient $c$ was found from numerical calculations to be equal to $c\!=\!0.823$~\cite{Beaul1997}. The same value for the numeric factor was used in~\cite{Tabut2006} to match experimentally measured drag coefficient on a sphere. The correction factor $X(n)$ was determined experimentally in~\cite{Gu1985}. Here we use polynomial fitting proposed in~\cite{Betan2015}
\begin{equation}\label{Xn}
X(n)=-1.1492\, n^2 + 0.8734 \,n + 1.2778.
\end{equation}
Combination of~(\ref{CD})--(\ref{Xn}) allows us to calculate theterminal settling velocity as
\begin{equation}\label{vs}
 v_z^p=\Bigl(\dfrac{(\rho^p\!-\!\rho^f)g d^{1+n}}{18 k X(n) }-c \,\dfrac{\tau_0 d^n}{k}\Bigr)^{1/n},
\end{equation}
whereby the settling velocity is zero if the expression in the parentheses is negative. The latter situation corresponds to the case, in which the gravitational force is not sufficient to exceed the yield stress of the carrying fluid. 

\subsection*{Filtration of Herschel-Bulkley fluid through proppant plug}

Filtration of a Herschel-Bulkley fluid through packed proppant plug was studied in~\cite{Ouyang2013}. Authors employed the capillary bundle model approach, in which the packed proppant is modeled by a series of capillary tubes with radius $R$. Fluid filtration velocity for such a problem is calculated as
\begin{equation}\label{vfilt}
\bvf= \Bigl(3+\dfrac{1}{n}\Bigr)^{-1} \Bigl(\dfrac{R}{2 k} |\!-\!\nabla p^f\!+\!\rho^f\bg| -\dfrac{\tau_0}{k} \Bigr)^{1/n} R \, \dfrac{\!-\!\nabla p^f\!+\!\rho^f\bg}{|\!-\!\nabla p^f\!+\!\rho^f\bg|},
\end{equation}
where $\bvf$ is the average fluid velocity and $R\!\approx \!0.1d$ is the effective radius whose value is calculated by comparing the analytical modeling results to numerical calculations. Note that authors in~\cite{Ouyang2013} used a more sophisticated relation between $R$ and $d$ that accounts for the variation with $n$ and the yield stress. Such a precise behavior is replaced by a typical value $R\!\approx \!0.1d$ since equation~(\ref{vfilt}) is used predominantly for estimation purposes.

\subsection*{Calculation of the viscous drag force}

To calculate the volumetric viscous force between fluid and particles, let us first consider the situation of particle settling for small concentrations. In this case, equations~(\ref{govp}), (\ref{govf}), and~(\ref{bf}) can be reduced to
\begin{eqnarray}\label{govsett}
0&=&\phi\rhop \bg+\bff,\notag\\
0&=&-\nabla p^f+(1\!-\!\phi)\rhof \bg-\bff,\\
\bff&=& (1\!-\!\phi)\phi\boldsymbol{f_v}-\phi \nabla p^f.\notag
\end{eqnarray} 
Solution of the above system of equations is
\begin{equation}\label{govsettsol}
\boldsymbol{f_v}= -(\rhop\!-\!\rhof)\bg,\qquad \nabla p^f=(\phi\rhop+(1\!-\!\phi)\rhof) \bg.
\end{equation} 
To ensure that the obtained force~(\ref{govsettsol}) leads to the settling velocity calculated in~(\ref{vs}), one should have
\begin{equation}\label{fv0}
\boldsymbol{f_{v,0}}=\dfrac{18 k X(n)}{d^{n+1}}\Bigl( |\bvf\!-\!\bvp|^n+c\dfrac{\tau_0 d^n}{k}\Bigr)\dfrac{\bvf\!-\!\bvp}{|\bvf\!-\!\bvp|},
\end{equation}
where the subscript ``0'' indicates that the expression for the force is applicable for small particle concentrations.

Let us now focus on the situation of fluid filtration through packed proppant, i.e. $\phi\!=\!\phi_{rcp}$. For this case, equations (\ref{govf}) and~(\ref{bf}) can be reduced to
\begin{eqnarray}\label{govfilt}
0&=&-\nabla p^f+(1\!-\!\phi)\rhof \bg-\bff,\notag\\
\bff&=& (1\!-\!\phi)\phi\boldsymbol{f_v}-\phi \nabla p^f.
\end{eqnarray} 
Solution of the above system of equations is
\begin{equation}\label{govfiltsol}
\boldsymbol{f_v}= \dfrac{1}{\phi}(- \nabla p^f\!+\!\rhof\bg).
\end{equation} 
Comparison between the latter result and~(\ref{vfilt}) gives
\begin{equation}\label{fv1}
\boldsymbol{f_{v,1}}=\dfrac{18 k X(n)}{ h_1 d^{n+1}}\Bigl(h_2^{-n} |\bvf\!-\!\bvp|^n+c\dfrac{\tau_0 d^n}{k}\Bigr)\dfrac{\bvf\!-\!\bvp}{|\bvf\!-\!\bvp|},\quad h_1= 9 c X(n)\dfrac{\phi R}{d},\quad h_2= c^{-1/n}\Bigl(3\!+\!\dfrac{1}{n}\Bigr)^{-1} \dfrac{R}{d},
\end{equation}
where the subscript ``1'' indicates that this force is applicable for large particle concentrations (i.e. packed proppant). It is worth noting that $h_1\approx0.55$ and $h_2\approx 0.03$ (for $n\!\gtrsim\! 0.3$) and the variation with $n$ is relatively mild. The following relations $R\!=\!0.1d$ and $\phi\!=\!\phi_{rcp}$ are used for the estimations.

To describe the viscous force in the whole range of particle concentrations, we combine~(\ref{fv0}) and~(\ref{fv1}) as
 \begin{equation}\label{fvfinal}
\boldsymbol{f_{v}}=\dfrac{18 k X(n) }{h_1(\phi) d^{n+1}}\Bigl(h_2(\phi)^{-n} |\bvf\!-\!\bvp|^n+c\dfrac{\tau_0 d^n}{k}\Bigr)\dfrac{\bvf\!-\!\bvp}{|\bvf\!-\!\bvp|},
\end{equation}
where $h_1(\phi)$ and $h_2(\phi)$ are the hindrance functions that depend only on particle concentration. 

For the case of Newtonian fluid, equation~(\ref{fvfinal}) reduces to
 \begin{equation}\label{fvfinalN}
\boldsymbol{f_{v}}=\dfrac{18 k (\bvf\!-\!\bvp)}{h_1(\phi) h_2(\phi)d^{2}},
\end{equation}
which can be used to calculate hindered settling velocity via~(\ref{govsettsol}) as
 \begin{equation}\label{settlN}
 \bvp\!-\!\bvf=\dfrac{h_1(\phi) h_2(\phi)d^{2}(\rho^p\!-\!\rho^f)\bg}{18 k}.
\end{equation}
Hindered settling is typically formulated in terms of the difference between the particle and mixture velocities, i.e.
\begin{equation}\label{settlNhind}
 \bvp\!-\!\bvs=\dfrac{h(\phi)d^{2}(\rho^p\!-\!\rho^f)\bg}{18 k},\qquad \bvs=\phi\bvp+(1\!-\!\phi)\bvf,
\end{equation}
where $h(\phi)$ is the hindrance function. Since $\bvp-\bvs=(1\!-\!\phi)( \bvp\!-\!\bvf)$, then equations~(\ref{settlN}) and~(\ref{settlNhind}) lead to
\begin{equation}\label{hindrelat}
h(\phi)=(1\!-\!\phi)h_1(\phi) h_2(\phi).
\end{equation}
A simple correlation $h(\phi)\!=\!(1\!-\!\phi)^\alpha$ was proposed in~\cite{Rich1954}, where $\alpha\!=\!5.1$ is a constant obtained by fitting the experimental data~\cite{Gars1977}. In view of equation~(\ref{hindrelat}), it is interesting to observe that the hindrance function $h(\phi)$ evaluated at $\phi\!=\!\phi_{rcp}$ for $\alpha\!=\!5.1$ agrees reasonably well with $(1\!-\!\phi)h_1 h_2$, where $h_1$ and $h_2$ are taken from~(\ref{fv1}). This indicates a consistency between the hindered settling results and permeability calculated in~\cite{Ouyang2013}. Given the characteristic values $h_1(\phi_{rcp})\!\approx\!0.55$ and $h_2(\phi_{rcp})\!\approx\! 0.03$, we define the functions $h_1(\phi)$ and $h_2(\phi)$ using~(\ref{hindrelat}) and $\alpha\!=\!5.1$ as
\begin{equation}\label{h1h2}
h_1(\phi)=(1\!-\!\phi)^{\alpha_1},\qquad h_2(\phi)=(1\!-\!\phi)^{\alpha_2}.
\end{equation}
where $\alpha_1\!=\!0.6$ and $\alpha_2\!=\!3.5$. Finally, the viscous force~(\ref{fvfinal}) is computed using expressions for the hindrance functions~(\ref{h1h2}) with the determined values of $\alpha_1$ and $\alpha_2$.

\section{Approximations for the functions $Q^s_1$, $Q^s_2$, $Q^p_1$, and $Q^p_2$}\label{QsQpappr}

To facilitate implementation of the developed model for slurry flow into numerical simulators, it is important to provide a quick way of computing the functions $Q^s_i$, $Q^p_i$ ($i\!=\!1,2$) introduced in~(\ref{QxQz}). In order to do this task, a database of numerical solutions $Q^s_i(\langle\phi\rangle)$, $Q^p_i(\langle\phi\rangle)$ ($i\!=\!1,2$) has been computed for the following ranges of the parameters
\begin{equation*}
0\leqslant\langle\phi\rangle\leqslant\phi_{rcp},\qquad 0.25\leqslant n\leqslant1,\qquad 0\leqslant T_0\leqslant0.75,\qquad 0\leqslant G_\rho\leqslant2,\qquad  0\leqslant\psi<\pi.
\end{equation*}
This range of parameters should be sufficient for most practical cases. Total of 400000 cases have been computed to uniformly sample the five-dimensional parametric space. Then, to provide a rapid way of computing the solution, an interpolating function is constructed, as outlined below.

As a starting point, we consider the following functional form for $Q^s_i$, $Q^p_i$ ($i\!=\!1,2$)
\begin{eqnarray}\label{Qapprox}
Q^s_1&=&\Bigl(1-\dfrac{\langle\phi\rangle}{\phi_{rcp}}\Bigr)^{p_1+p_2 {\langle\phi\rangle}/{\phi_{rcp}}+p_3 ({\langle\phi\rangle}/{\phi_{rcp}})^2+p_4 ({\langle\phi\rangle}/{\phi_{rcp}})^3 }\, Q_{HB},\notag\\
Q^p_1&=&\langle\phi\rangle\Bigl[1+\Bigl(p_5+p_6 \dfrac{\langle\phi\rangle}{\phi_{rcp}}\Bigr)\Bigl(1-\dfrac{\langle\phi\rangle}{\phi_{rcp}}\Bigr)\Bigr]\, Q^s_1,\notag\\
 Q^s_2&=&\langle\phi\rangle\Bigl(1-\dfrac{\langle\phi\rangle}{\phi_{rcp}}\Bigr)\, \Bigl(p_7+p_8 \dfrac{\langle\phi\rangle}{\phi_{rcp}}\Bigr)\, Q^s_1,\\
 Q^p_2&=&\langle\phi\rangle\Bigl[1+\Bigl(p_9+p_{10} \dfrac{\langle\phi\rangle}{\phi_{rcp}}\Bigr)\,\Bigl(1-\dfrac{\langle\phi\rangle}{\phi_{rcp}}\Bigr)\Bigr]\,Q^s_2,\notag
\end{eqnarray}  
where the coefficients $p_i$ ($i=1..10$) are the functions of $G_\rho$, $\cos\psi$, $n$, and $T_0$. To proceed with the approximation we perform polynomial fitting for the coefficients $p_i$ ($i=1..10$) as follows
\begin{equation}\label{Cijklm}
p_i = \sum_{j,k,l,m=1}^3 C_{ijklm} G_\rho^{j-1} \cos(\psi)^{k-1} T_0^{l-1} n^{m-1},
\end{equation}
where $C_{ijklm}$ ($i=1..10$, $j=1..3$, $k=1..3$, $l=1..3$, $m=1..3$) are constants, see Tables~\ref{tabC1}--\ref{tabC10}.

\begin{figure}[h]
\centerline{\includegraphics[width=0.99\linewidth]{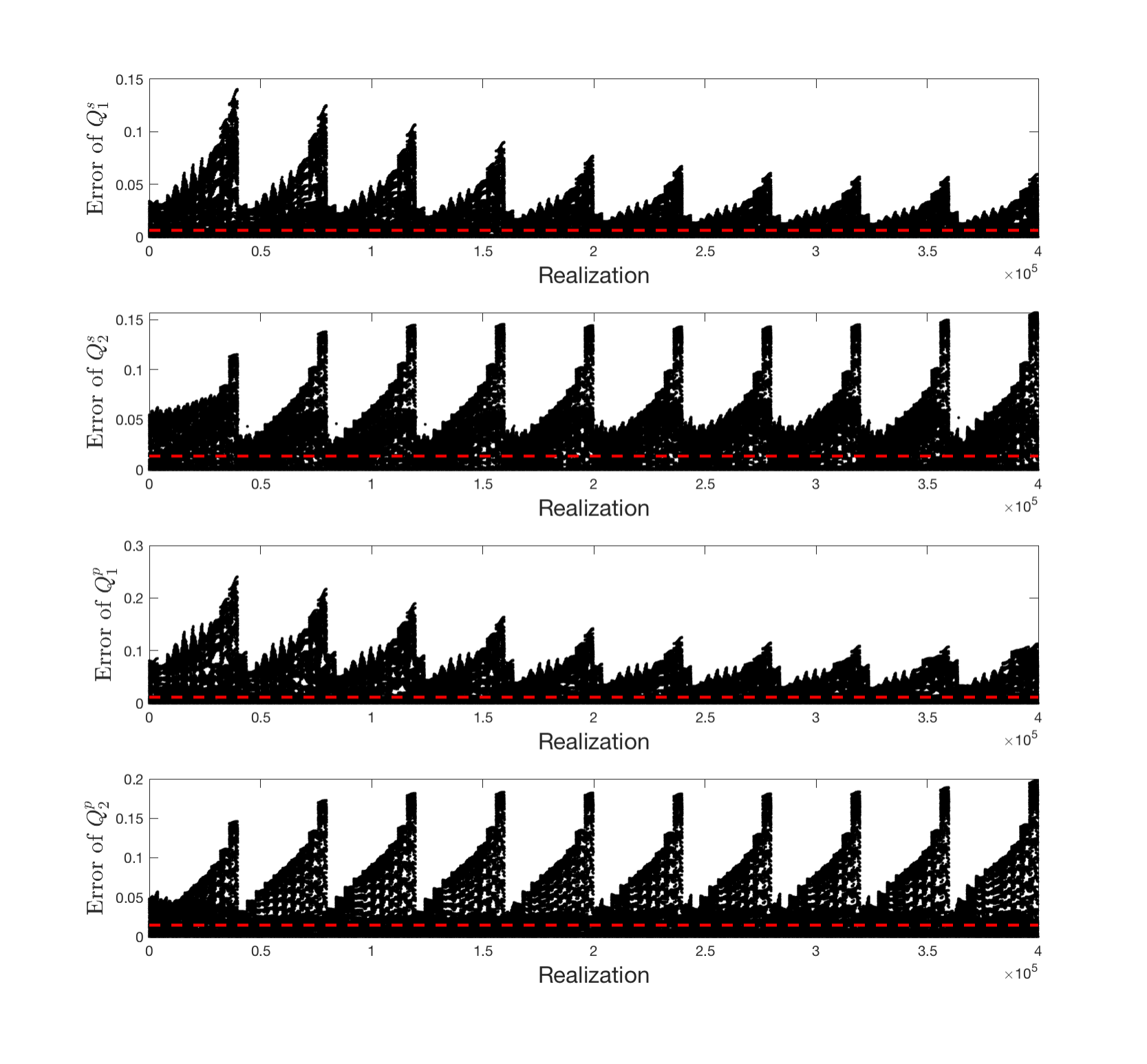}} 
\caption{Relative error of the approximations~(\ref{Qapprox}) versus realization in the parametric space. Dashed red lines show average error.}
 \label{fig_error}
\end{figure}

As an illustration, Fig.~\ref{fig_error} shows the relative error between the numerically computed functions  $Q^s_i$, $Q^p_i$ ($i\!=\!1,2$) and their approximations computed using~(\ref{Qapprox}), (\ref{Cijklm}) versus realization number in the database of runs. Dashed red lines indicate average error of the approximation. One can see that the developed approximation is able to reasonably accurately capture behavior of the solution in the whole parametric space. There are some points that have larger level of error. These are associated with large values of $G_\rho$, see complex variation with $G_\rho$ in Figs.~\ref{figQs} and~\ref{figQp}.

\begin{table}
\caption{Fitting coefficients in~(\ref{Cijklm}) for $i\!=\!1$.}
\begin{center}
\begin{tabular}{c||ccccccccc}
$i\!=\!1$ & $lm\!=\!11$ &    $lm\!=\!21$ &   $lm\!=\!31$ &    $lm\!=\!12$ &   $lm\!=\!22$ &    $lm\!=\!32$ &   $lm\!=\!13$ &    $lm\!=\!23$ &   $lm\!=\!33$ \\
\hline \hline
 $jk\!=\!11$  & 0.0081&    1.1311&   -5.5492&    1.7919&   -2.0709&    8.7765&   -0.9453&    1.1044&   -4.8743\\
 $jk\!=\!21$ &-0.0213&    0.3317&   -0.4174&    0.0378&   -0.5948&    0.7734&   -0.0216&    0.3384&   -0.4423\\
 $jk\!=\!31$  & 0.0255&  -0.4741&    0.6929&   -0.0392&    0.7862&  -1.1600&    0.0233&   -0.4448&    0.6541\\
 $jk\!=\!12$ &0.0011&   -0.0629&    0.1519&    0.0054&    0.0988&   -0.2442&   -0.0023&   -0.0581&    0.1399\\
 $jk\!=\!22$  & 0.2381&    0.6468&   -0.5650&  -0.5216&   -0.3666&    0.7779&    0.2109&    0.3033&   -0.4500\\
 $jk\!=\!32$  & 0.0295&   -0.4273&    0.5297&   -0.0257&    0.6712&  -0.8089&    0.0121&   -0.3654&   0.4431\\
 $jk\!=\!13$  &-0.0202&    0.3405&   -0.4487&    0.0475&   -0.5745&    0.7536&   -0.0256&    0.3211&   -0.4238\\
 $jk\!=\!23$  & 0.2493&  -3.9146&    5.7842&   -0.4929&    6.4178&   -9.4354&   0.2704&   -3.5762&    5.2652\\
 $jk\!=\!33$  &-0.1043&   1.9678&   -2.7760&    0.1804&   -3.1777&    4.5234&   -0.1073&    1.7928&   -2.5400
\end{tabular}
\end{center} 
\label{tabC1}
\end{table}

\begin{table}
\caption{Fitting coefficients in~(\ref{Cijklm}) for $i\!=\!2$.}
\begin{center}
\begin{tabular}{c||ccccccccc}
$i\!=\!2$ & $lm\!=\!11$ &    $lm\!=\!21$ &   $lm\!=\!31$ &    $lm\!=\!12$ &   $lm\!=\!22$ &    $lm\!=\!32$ &   $lm\!=\!13$ &    $lm\!=\!23$ &   $lm\!=\!33$ \\
\hline \hline
  $jk\!=\!11$  &   5.2865 &  -9.4313 &  36.6969  & -7.9636  & 14.2911 & -57.5183  &  4.3865  & -7.9763 &  32.0268\\
  $jk\!=\!21$ & 0.3344 &  -4.8720  &  6.3871 &  -0.7567  &  8.9324 & -11.8727 &   0.3975 &  -4.9977  &  6.7222\\
  $jk\!=\!31$ &-0.4015 &   7.1897 & -10.9417  &  0.9121 & -12.1824 &  18.4080  & -0.4781 &   6.7462  &-10.2643\\
  $jk\!=\!12$ &-0.0161  &  1.0687  & -2.4719 &  -0.0538  & -1.6753  &  3.9682  &  0.0145   & 0.9966   &-2.2846\\
  $jk\!=\!22$  &  -0.1479  & -9.5915  & 11.0536  &  0.4545  & 13.5067 & -16.3528  & -0.0310 &  -8.0811  &  9.5001\\
  $jk\!=\!32$  &  -0.3154 &   6.3837  & -7.7220 &   0.0629  & -9.5862  & 11.6220 &  -0.0890  &  5.4579  & -6.5506\\
  $jk\!=\!13$  &  0.3234 &  -4.9447 &   6.5064  & -0.6957 &   8.3770 & -10.9602  &  0.3673  & -4.6530   & 6.1376\\
  $jk\!=\!23$  &-3.9508  & 58.2192 & -86.4337  &  7.4737 & -95.7264 & 141.1272 &  -4.0301  & 53.0954  &-78.5418\\
  $jk\!=\!33$  & 1.8520 & -29.6642 &  43.1362  & -3.8759  & 49.1134 & -70.7919  &  2.0345 & -27.0785  & 39.2705
\end{tabular}
\end{center} 
\label{tabC2}
\end{table}

\begin{table}
\caption{Fitting coefficients in~(\ref{Cijklm}) for $i\!=\!3$.}
\begin{center}
\begin{tabular}{c||ccccccccc}
$i\!=\!3$ & $lm\!=\!11$ &    $lm\!=\!21$ &   $lm\!=\!31$ &    $lm\!=\!12$ &   $lm\!=\!22$ &    $lm\!=\!32$ &   $lm\!=\!13$ &    $lm\!=\!23$ &   $lm\!=\!33$ \\
\hline \hline
 $jk\!=\!11$  & -13.9649 &  34.3294& -118.5802 &  21.5559 & -52.3130  &185.8810 & -11.9135  & 29.3718 &-103.6740\\
 $jk\!=\!21$ &  -1.0457 &  15.1503&  -19.9370 &   2.3225 & -27.8562  & 37.2858   &-1.2325   &15.6307  &-21.1562\\
 $jk\!=\!31$ & 1.4025 & -23.0211&   35.3626  & -2.9719  & 38.9161 & -59.4169   & 1.5735  &-21.5861  & 33.1536\\
 $jk\!=\!12$ & 0.0952  & -3.8939  &  8.6794  &  0.0177  &  6.1898 & -13.9679   & 0.0293   &-3.6266  &  8.0023\\
 $jk\!=\!22$  & -1.2551 &  31.5944 & -37.7972 &   1.5539 & -47.7772  & 57.3433  & -1.1377  & 27.4676 & -32.6502\\
 $jk\!=\!32$  &  1.0215 & -20.9103 &  25.7678  & -0.5348  & 31.5552 & -38.9634  &  0.4529  &-17.9014  & 21.9020\\
 $jk\!=\!13$  & -1.0321 &  15.3906 & -20.1948  &  2.0696 & -26.0249  & 34.0688  & -1.1060  & 14.4988  &-19.1158\\
 $jk\!=\!23$  &12.8485 &-185.9555 & 277.1381 & -23.2731 & 305.2181& -452.3869 &  12.6351 &-169.5155 & 251.9373\\
 $jk\!=\!33$  & -6.2751 &  94.8873 &-139.0668  & 12.3939 &-156.6903 & 227.8614  & -6.5514   &86.4987 &-126.4695
\end{tabular}
\end{center} 
\label{tabC3}
\end{table}

\begin{table}
\caption{Fitting coefficients in~(\ref{Cijklm}) for $i\!=\!4$.}
\begin{center}
\begin{tabular}{c||ccccccccc}
$i\!=\!4$ & $lm\!=\!11$ &    $lm\!=\!21$ &   $lm\!=\!31$ &    $lm\!=\!12$ &   $lm\!=\!22$ &    $lm\!=\!32$ &   $lm\!=\!13$ &    $lm\!=\!23$ &   $lm\!=\!33$ \\
\hline \hline
  $jk\!=\!11$  &  13.1587&  -36.9251&  121.7656 & -20.6975 &  56.8422 &-191.4346  & 11.4218  &-31.8553 & 106.7031\\
  $jk\!=\!21$ &   0.8609  &-12.4475  & 16.4058 &  -1.8545 &  22.9422  &-30.8895   & 0.9980  &-12.9239   &17.5771\\
  $jk\!=\!31$ &  -1.2323  & 19.4554 & -29.9663  &  2.4598  &-32.8026  & 50.3278  & -1.3195  & 18.2447  &-28.1202\\
  $jk\!=\!12$ &  -0.1222 &   3.6996  & -7.9810  &  0.0968   &-5.9326  & 12.8676   &-0.0800   & 3.4383  & -7.3437\\
  $jk\!=\!22$  & 1.2902 & -26.8707 &  31.0634  & -1.6513  & 41.2270  &-47.4592  &  1.0540  &-23.3543  & 26.7717\\
  $jk\!=\!32$  & -0.9295  & 18.1411 & -22.7626  &  0.7769  &-27.5817  & 34.5895   &-0.5344  & 15.5556  &-19.3682\\
  $jk\!=\!13$  &   0.8625 & -12.6859  & 16.6514  & -1.6534  & 21.4279  &-28.1341  &  0.8941  &-11.9725  & 15.8182\\
  $jk\!=\!23$  &-10.9810 & 157.3072 &-235.6724  & 19.3319 &-257.8034 & 384.5904  &-10.5655  &143.3760 &-214.3361\\
  $jk\!=\!33$  & 5.4506 & -80.2985  &118.2873 & -10.2484  &132.1549 &-193.5573  &  5.4820 & -73.1400  &107.5671
\end{tabular}
\end{center} 
\label{tabC4}
\end{table}

\begin{table}
\caption{Fitting coefficients in~(\ref{Cijklm}) for $i\!=\!5$, multiplied by $10^3$.}
\begin{center}
\begin{tabular}{c||ccccccccc}
$i\!=\!5$ & $lm\!=\!11$ &    $lm\!=\!21$ &   $lm\!=\!31$ &    $lm\!=\!12$ &   $lm\!=\!22$ &    $lm\!=\!32$ &   $lm\!=\!13$ &    $lm\!=\!23$ &   $lm\!=\!33$ \\
\hline \hline
 $jk\!=\!11$  &  23.7778 &  -6.3481 & -15.6413 & 295.7701 &-309.8799 &  23.6318& -119.3150 & 131.3341  &-21.0767\\
 $jk\!=\!21$ &    -0.0362  &  0.7032 &  -0.8565  &  4.4127 &  -8.8737  &  4.5449&   -0.8743  & -0.1347  &  1.6801\\
 $jk\!=\!31$ & -0.1377 &  -0.6763 &   0.8632 &  -8.5384  & 16.0936  & -8.9531 &   2.2870  & -2.0915  & -0.3415\\
 $jk\!=\!12$ &   -0.0023 &   0.0386 &  -0.0772 &   0.5935 &  -1.8808  &  1.2395  & -0.0221  & -0.2532  &  0.5312\\
 $jk\!=\!22$  &  -1.4992 &   3.6448 &  -1.7126  & 31.6232 & -79.2639 &  55.5472 & -10.3219  & 27.7229 & -21.4668\\
 $jk\!=\!32$  & -0.0754  &  2.4666   &-3.1637  &  6.8282  & -7.4291  & -1.4536   &-0.9324 &  -5.1497  &  8.6280\\
 $jk\!=\!13$  &   -0.0342  & -0.0154  &  0.0900  &  0.9525 &  -4.0735  &  3.8704  &  0.0015  & -0.0559  &  0.1129\\
 $jk\!=\!23$  &   0.1252  &  1.4010 &  -2.7770 &  -8.1276 &  37.9593  &-41.5991 &   0.6196 &  -3.9509  &  5.8349\\
 $jk\!=\!33$  &  0.1627  &  1.5212  & -1.4539  & 24.8579 & -57.3640  & 40.0783  & -5.6849  &  5.9328  & -0.2936
\end{tabular}
\end{center} 
\label{tabC5}
\end{table}

\begin{table}
\caption{Fitting coefficients in~(\ref{Cijklm}) for $i\!=\!6$, multiplied by $10^3$.}
\begin{center}
\begin{tabular}{c||ccccccccc}
$i\!=\!6$ & $lm\!=\!11$ &    $lm\!=\!21$ &   $lm\!=\!31$ &    $lm\!=\!12$ &   $lm\!=\!22$ &    $lm\!=\!32$ &   $lm\!=\!13$ &    $lm\!=\!23$ &   $lm\!=\!33$ \\
\hline \hline
  $jk\!=\!11$  &    66.9886 &-143.8778   &78.5138  &  7.8389  &179.4146 &-262.4848 & -28.1583  &-17.4663  & 76.3498\\
  $jk\!=\!21$ &    0.0378  & -0.9513 &   1.3479  & -5.3565  & 11.2493 &  -5.2237    &1.0534   &-0.0573   &-2.1174\\
  $jk\!=\!31$ & -0.3046  &  0.8083   &-0.3878   & 8.3957 & -20.7632  & 15.1431   &-1.9375  &  3.1182  & -1.4735\\
  $jk\!=\!12$ & -0.0400  &  0.2085   &-0.1041   &-0.7896   & 3.4356   &-2.3102    &0.0863  & -0.2588  & -0.2617\\
  $jk\!=\!22$  & -0.7999  &  2.7632  & -2.9273  &-13.6109 &  60.8601 & -64.3815   & 2.5442  &-16.6725  & 20.9005\\
  $jk\!=\!32$  &  -0.0020  & -4.8504 &   6.7561 &  -8.0120 &  -4.0980  & 20.4186  &  1.1200  & 11.6557 & -18.3863\\
  $jk\!=\!13$  &  0.0199   & 0.5596  & -0.7810  & -1.0572  &  8.2794  & -9.2356  & -0.0043   &-1.5300   & 1.9721\\
  $jk\!=\!23$  &  0.1731  & -8.1584  & 12.6004 &  10.7591 & -87.4057  &112.5521 &  -1.3324  & 23.4003 & -34.5606\\
  $jk\!=\!33$  &   -0.2119  &  2.8915  & -4.6911  &-27.7437 &  84.7213 & -74.2583 &   6.1877  &-14.9904   &12.5086
\end{tabular}
\end{center} 
\label{tabC6}
\end{table}

\begin{table}
\caption{Fitting coefficients in~(\ref{Cijklm}) for $i\!=\!7$, multiplied by $10^3$.}
\begin{center}
\begin{tabular}{c||ccccccccc}
$i\!=\!7$ & $lm\!=\!11$ &    $lm\!=\!21$ &   $lm\!=\!31$ &    $lm\!=\!12$ &   $lm\!=\!22$ &    $lm\!=\!32$ &   $lm\!=\!13$ &    $lm\!=\!23$ &   $lm\!=\!33$ \\
\hline \hline
  $jk\!=\!11$  &23.8579 &  -6.9152  &-16.1497  &294.8509 &-308.5847  & 23.2499 &-118.9434 & 130.7795 & -20.8737\\
  $jk\!=\!21$ & -0.0480  &  0.6830 &  -0.8450   & 4.3899   &-8.8294   & 4.5213  & -0.8702  & -0.1320   & 1.6704\\
  $jk\!=\!31$ &  -0.1178 &  -0.6371 &   0.8144  & -8.4941  & 16.0130  & -8.9130  &  2.2761  & -2.0848  & -0.3347\\
  $jk\!=\!12$ & -0.0024  &  0.0249  & -0.0678  &  0.5909  & -1.8749   & 1.2358  & -0.0217 &  -0.2528  &  0.5299\\
  $jk\!=\!22$  & -1.5295 &   3.5478  & -1.5080 &  31.4846 & -79.0056  & 55.4203  &-10.2770 &  27.6484  &-21.4352\\
  $jk\!=\!32$  &-0.1010  &  2.5760   &-3.3341  &  6.7933  & -7.3611   &-1.4904  & -0.9260   &-5.1448   & 8.6121\\
  $jk\!=\!13$  & -0.0366  & -0.0538  &  0.1380  &  0.9479  & -4.0635   & 3.8642  &  0.0022  & -0.0550  &  0.1105\\
  $jk\!=\!23$  &   0.1395 &   1.7787 &  -3.3098 &  -8.0908 &  37.8727  &-41.5377 &   0.6131  & -3.9522  &  5.8452\\
  $jk\!=\!33$  &  0.0903   & 1.2965  & -1.1316  & 24.7343  &-57.1255  & 39.9478  & -5.6568  &  5.9203 &  -0.3169
\end{tabular}
\end{center} 
\label{tabC7}
\end{table}

\begin{table}
\caption{Fitting coefficients in~(\ref{Cijklm}) for $i\!=\!8$, multiplied by $10^3$.}
\begin{center}
\begin{tabular}{c||ccccccccc}
$i\!=\!8$ & $lm\!=\!11$ &    $lm\!=\!21$ &   $lm\!=\!31$ &    $lm\!=\!12$ &   $lm\!=\!22$ &    $lm\!=\!32$ &   $lm\!=\!13$ &    $lm\!=\!23$ &   $lm\!=\!33$ \\
\hline \hline
   $jk\!=\!11$  & 62.2927 &-142.0252  & 80.1815  &  9.0309  &177.0034 &-261.1947 & -28.6132  &-16.4997  & 75.7845\\
   $jk\!=\!21$ &  0.0475  & -0.9020 &   1.3288  & -5.3284&   11.1977  & -5.1993   & 1.0485  & -0.0625  & -2.1027\\
   $jk\!=\!31$ &  -0.3031  &  0.6945 &  -0.2454  &  8.3458 & -20.6630 &  15.0881 &  -1.9268  &  3.1127  & -1.4822\\
   $jk\!=\!12$ &   -0.0449 &   0.2501 &  -0.1325  & -0.7873  &  3.4296 &  -2.3061  &  0.0860 &  -0.2591   &-0.2605\\
   $jk\!=\!22$  &   -1.0244   & 3.0554 &  -3.5742 & -13.3927  & 60.4767  &-64.2112  &  2.4729  &-16.5689 &  20.8701\\
   $jk\!=\!32$  &  0.0287 &  -5.2043 &   7.3001  & -7.9718  & -4.1704  & 20.4532  &  1.1134  & 11.6442 & -18.3610\\
   $jk\!=\!13$  &  0.0185  &  0.6739  & -0.9238 &  -1.0532  &  8.2699  & -9.2296  & -0.0048  & -1.5311   & 1.9746\\
   $jk\!=\!23$  &    0.1952 &  -9.3219  & 14.2553   &10.7217  &-87.2982  &112.4565   &-1.3260 &  23.3945  &-34.5583\\
   $jk\!=\!33$  &-0.1686 &   3.5580  & -5.6558  &-27.5983  & 84.4213 & -74.0804   & 6.1578  &-14.9784  & 12.5354
\end{tabular}
\end{center} 
\label{tabC8}
\end{table}

\begin{table}
\caption{Fitting coefficients in~(\ref{Cijklm}) for $i\!=\!9$, multiplied by $10^2$.}
\begin{center}
\begin{tabular}{c||ccccccccc}
$i\!=\!9$ & $lm\!=\!11$ &    $lm\!=\!21$ &   $lm\!=\!31$ &    $lm\!=\!12$ &   $lm\!=\!22$ &    $lm\!=\!32$ &   $lm\!=\!13$ &    $lm\!=\!23$ &   $lm\!=\!33$ \\
\hline \hline
  $jk\!=\!11$  &  3.9230  &  3.9767 &  -8.2924  & 75.1871 &-129.4272 &  60.9150 & -29.2629  & 44.9833 & -17.7408\\
  $jk\!=\!21$ & -0.0423  &  0.3367  & -0.3725  &  1.1531   &-3.1431  &  2.2818   &-0.2812 &   0.3564  &  0.0215\\
  $jk\!=\!31$ &   0.0547 &  -0.5498 &   0.5867  & -2.4011  &  6.1307 &  -4.4214  &  0.6160  & -0.8220   & 0.1357\\
  $jk\!=\!12$ &   0.0071  & -0.0085 &  -0.0080 &   0.0661 &  -0.2732  &  0.2185  & -0.0319 &   0.1009  & -0.0604\\
  $jk\!=\!22$  &  -0.6490 &   2.4581 &  -2.0754  &  8.7872 & -26.6442  & 21.2240  & -2.2818  &  5.5477   &-3.7038\\
  $jk\!=\!32$  & -0.0458 &   0.7559  & -0.9439  &  1.5682  & -3.3006  &  1.8202   &-0.3159  & -0.4077  &  1.1363\\
  $jk\!=\!13$  & 0.0061   &-0.0347  &  0.0375  &  0.1036   &-0.6027  &  0.6396  & -0.0523   & 0.2474   &-0.2385\\
  $jk\!=\!23$  &   -0.0331   & 0.3803 &  -0.5914  & -1.2111  &  6.5215 &  -7.4122  &  0.4357   &-2.1563  &  2.4185\\
  $jk\!=\!33$  & -0.1618  &  1.3920  & -1.4318  &  6.4513  &-18.3163  & 14.4596  & -1.5388  &  2.2589  & -0.6236
\end{tabular}
\end{center} 
\label{tabC9}
\end{table}

\begin{table}
\caption{Fitting coefficients in~(\ref{Cijklm}) for $i\!=\!10$, multiplied by $10^2$.}
\begin{center}
\begin{tabular}{c||ccccccccc}
$i\!=\!10$ & $lm\!=\!11$ &    $lm\!=\!21$ &   $lm\!=\!31$ &    $lm\!=\!12$ &   $lm\!=\!22$ &    $lm\!=\!32$ &   $lm\!=\!13$ &    $lm\!=\!23$ &   $lm\!=\!33$ \\
\hline \hline
  $jk\!=\!11$  &14.8085 & -34.6195 &  21.0900 & -31.8899 & 124.0410 &-113.3544   & 6.2874  &-30.0723  & 30.3150\\
  $jk\!=\!21$ & 0.0513  & -0.3988  &  0.4573  & -1.4293  &  3.9358   &-2.7231   & 0.3448   &-0.4505  & -0.0806\\
  $jk\!=\!31$ & -0.1608   & 0.6456  & -0.5487  &  2.6383  & -7.6556  &  6.0081  & -0.5988   & 0.9909  & -0.3509\\
  $jk\!=\!12$ &   -0.0208  &  0.0801 &  -0.0483  & -0.0648  &  0.4186  & -0.3153  &  0.0464   &-0.2210  &  0.1513\\
  $jk\!=\!22$  &   0.5214  & -1.9767 &   1.6631 &  -7.4051  & 26.8915 & -24.4881  &  1.1203  & -3.2785  &  2.8035\\
  $jk\!=\!32$  & 0.0336  & -1.2899   & 1.7385   &-1.8332   & 1.8336   & 0.8175   & 0.3686    &1.4439   &-2.7364\\
  $jk\!=\!13$  &   -0.0178   & 0.1936  & -0.2302 &  -0.0603 &   1.0570  & -1.2956  &  0.0589  & -0.5545  &  0.6260\\
  $jk\!=\!23$  &  0.1450&   -2.0562  &  2.9382  &  1.3189 & -13.2383  & 17.5854 &  -0.6025 &   5.6536  & -7.4423\\
  $jk\!=\!33$  & 0.2073  & -0.7181  &  0.3798  & -7.4959  & 24.4488 & -21.2568  &  1.6958  & -3.6708   & 2.3604
\end{tabular}
\end{center} 
\label{tabC10}
\end{table}


\begin{thebibliography}{10}

\bibitem{Econo2000}
M.J. Economides and K.G. Nolte, editors.
\newblock {\em Reservoir Stimulation}.
\newblock John Wiley \& Sons, Chichester, UK, 3rd edition, 2000.

\bibitem{King2012}
G.~King.
\newblock Hydraulic fracturing 101: What every representative,
  environmentalist, regulator,reporter, investor, university researcher,
  neighbor and engineer should know about estimatingfrac risk and improving
  frac performance in unconventional gas and oil wells.
\newblock In {\em In Proceedings of SPE Hydraulic Fracturing Technology
  Conference and Exhibition, 6-8 February, The Woodlands, Texas, USA, SPE
  152596}, 2012.

\bibitem{Dane1978}
A.A. Daneshy.
\newblock Numerical solution of sand transport in hydraulic fracturing.
\newblock {\em J. Pet. Tech.}, pages 132--140, 1978.

\bibitem{Pear1994}
J.R.A. Pearson.
\newblock On suspension transport in a fracture: framework for a global model.
\newblock {\em J. Non-Newtonian Fluid. Mech.}, 54:503--512, 1994.

\bibitem{Mobbs2001}
A.T. Mobbs and P.S. Hammond.
\newblock Computer simulations of proppant transport in a hydraulic fracture.
\newblock {\em SPE Prod. Facil.}, pages 112--121, 2001.

\bibitem{Shokir2007}
E.M. Shokir and A.A. Al-Quraishi.
\newblock Experimental and numerical investigation of proppant placement in
  hydraulic fractures.
\newblock In {\em SPE Latin American and Caribbean Petroleum Engineering
  Conference}, 2007.

\bibitem{Eskin2008}
D.~Eskin and M.J. Miller.
\newblock A model of non-newtonian slurry flow in a fracture.
\newblock {\em Powder Technology}, 182:313--322, 2008.

\bibitem{Boro2010}
S.A. Boronin and A.A. Osiptsov.
\newblock Two-continua model of suspension flow in a hydraulic fracture.
\newblock {\em Doklady Physics}, 55:199--202, 2010.

\bibitem{Dont2014b}
E.V. Dontsov and A.P. Peirce.
\newblock Slurry flow, gravitational settling, and a proppant transport model
  for hydraulic fractures.
\newblock {\em J. Fluid Mech.}, 760:567--590, 2014.

\bibitem{Lecam2014}
B.~Lecampion and D.~Garagash.
\newblock Confined flow of suspensions modeled by a frictional rheology.
\newblock {\em J. Fluid Mech.}, 759:197--235, 2014.

\bibitem{Shioz2016}
S.~Shiozawa and M.~McClure.
\newblock Simulation of proppant transport with gravitational settling and
  fracture closure in a three-dimensional hydraulic fracturing simulator.
\newblock {\em J. Petr. Sci. and Eng.}, 138:298--314, 2016.

\bibitem{Donts2019}
E.V. Dontsov, S.A. Boronin, A.A. Osiptsov, and D.Yu. Derbyshev.
\newblock Lubrication model of suspension flow in a hydraulic fracture with
  frictional rheology for shear-induced migration and jamming.
\newblock {\em Proc. R. Soc. A.}, 475:20190039, 2019.

\bibitem{Isaev2023}
V.I. Isaev, S.V. Idimeshev, L.G. Semin, and A.A. Tikhonov.
\newblock A lagrangian method for slurry flow modeling in hydraulic fractures.
\newblock {\em Geoenergy Science and Engineering}, 231:212272, 2023.

\bibitem{Adachi2007}
J.~Adachi, E.~Siebrits, A.~Peirce, and J.~Desroches.
\newblock Computer simulation of hydraulic fractures.
\newblock {\em Int. J. Rock Mech. Min. Sci.}, 44:739--757, 2007.

\bibitem{Osipts2017}
A.A. Osiptsov.
\newblock Fluid mechanics of hydraulic fracturing: a review.
\newblock {\em J. Petr. Sci. and Eng.}, 156:513--535, 2017.

\bibitem{Leigh1987}
D.~Leighton and A.~Acrivos.
\newblock The shear-induced migration of particles in concentrated suspensions.
\newblock {\em J. Fluid Mech.}, 181:415--439, 1987.

\bibitem{Nott1994}
P.R. Nott and J.F. Brady.
\newblock Pressure-driven flow of suspensions: simulation and theory.
\newblock {\em J. Fluid Mech.}, 275:157--199, 1994.

\bibitem{Phil1992}
R.J. Phillips, R.~C. Armstrong, R.~A. Brown, A.~Graham, and J.~R. Abbott.
\newblock A constitutive model for concentrated suspensions that accounts for
  shear-induced particle migration.
\newblock {\em Phys. Fluids A}, 4:30--40, 1992.

\bibitem{Nott2011}
P.~R. Nott, E.~Guazzelli, and O.~Pouliquen.
\newblock The suspension balance model revisited.
\newblock {\em Phys. Fluids}, 23:043304, 2011.

\bibitem{Morr1999}
J.F. Morris and F.~Boulay.
\newblock Curvilinear flows of noncolloidal suspensions: The role of normal
  stresses.
\newblock {\em J. Rheol.}, 43:1213--1237, 1999.

\bibitem{Eins1905}
A.~Einstein.
\newblock Über die von der molekularkinetischen theorie der wärme geforderte
  bewegung von in ruhenden flüssigkeiten suspendierten teilchen (german).
\newblock {\em Ann. Phys.}, 322:549--560, 1905.

\bibitem{Mill2006}
R.M. Miller and J.F. Morris.
\newblock Normal stress-driven migration and axial development in
  pressure-driven flow of concentrated suspensions.
\newblock {\em J. Non-Newtonian Fluid. Mech.}, 135:149--165, 2006.

\bibitem{Boyer2011}
F.~Boyer, E.~Guazzelli, and O.~Pouliquen.
\newblock Unifying suspension and granular rheology.
\newblock {\em Phys. Rev. Lett.}, 107:188301, 2011.

\bibitem{Dont2015}
E.V. Dontsov and A.P. Peirce.
\newblock Proppant transport in hydraulic fracturing: {C}rack tip screen-out in
  {KGD} and {P3D} models.
\newblock {\em Int. J. Solids Struct}, 63:206--218, 2015.

\bibitem{Dont2015b}
E.V. Dontsov and A.P. Peirce.
\newblock A lagrangian approach to modelling proppant transport with tip
  screen-out in kgd hydraulic fractures.
\newblock {\em Rock Mech Rock Eng}, 48:2541–2550, 2015.

\bibitem{Wang2018}
Jiehao Wang, Derek Elsworth, and Martin~K. Denison.
\newblock Propagation, proppant transport and the evolution of transport
  properties of hydraulic fractures.
\newblock {\em Journal of Fluid Mechanics}, 855:503–534, 2018.

\bibitem{Wang2018b}
Jiehao Wang and Derek Elsworth.
\newblock Role of proppant distribution on the evolution of hydraulic fracture
  conductivity.
\newblock {\em Journal of Petroleum Science and Engineering}, 166:249--262,
  2018.

\bibitem{Yang2024}
Peng Yang, Shicheng Zhang, Yushi Zou, Anhai Zhong, Feng Yang, Danyang Zhu, and
  Ming Chen.
\newblock Numerical simulation of integrated three-dimensional hydraulic
  fracture propagation and proppant transport in multi-well pad fracturing.
\newblock {\em Computers and Geotechnics}, 167:106075, 2024.

\bibitem{Oh2015}
S.~Oh, Y.~q.~Song, D.~Garagash, B.~Lecampion, and J.~Desroches.
\newblock Pressure-driven suspension flow near jamming.
\newblock {\em Phys. Rev. Lett.}, 114:088301, 2015.

\bibitem{Dago2015}
S.~Dagois-Bohy, S.~Hormozi, E.~Guazzelli, and O.~Pouliquen.
\newblock Rheology of dense suspensions of non-colloidal spheres in
  yield-stress fluids.
\newblock {\em J. Fluid Mech.}, 776:R2, 2015.

\bibitem{Chat2008}
X.~Chateau, G.~Ovarlez, and K.L. Trung.
\newblock Homogenization approach to the behavior of suspensions of
  noncolloidal particles in yield stress fluids.
\newblock {\em J. Rheol.}, 52:489--506, 2008.

\bibitem{Tapia2017}
F.~Tapia, S.~Shaikh, J.E. Butler, O.~Pouliquen, and E.~Guazzelli.
\newblock Rheology of concentrated suspensions of non-colloidal rigid fibers.
\newblock {\em J. Fluid Mech.}, 827:R5, 2017.

\bibitem{BoyerPHD}
F.~Boyer.
\newblock {\em Suspensions concentr\'{e}es: exp\'{e}riences originales de
  rh\'{e}ologie}.
\newblock PhD thesis, Aix-Marseille Universit\'{e}, 2011.

\bibitem{Dont2014}
E.V. Dontsov and A.P. Peirce.
\newblock A new technique for proppant schedule design.
\newblock {\em Hydraulic Fracturing Journal}, 1(3), 2014.

\bibitem{Ansley1967}
R.~W. Ansley and T.~N. Smith.
\newblock Motion of spherical particles in a {B}ingham plastic.
\newblock {\em AIChE J.}, 13:1193--1196, 1967.

\bibitem{Beaul1997}
M.~Beaulne and E.~Mitsoulis.
\newblock Creeping motion of a sphere in tubes filled with {H}erschel-{B}ulkley
  fluids.
\newblock {\em J. Non-Newtonian Fluid Mech.}, 72:55--71, 1997.

\bibitem{Tabut2006}
H.~Tabuteau, P.~Coussot, and J.~R. de~Bruyn.
\newblock Drag force on a sphere in steady motion through a yield-stress fluid.
\newblock {\em J. Rheol.}, 51:125--137, 2007.

\bibitem{Gu1985}
D.~Gu and R.I. Tanner.
\newblock The drag on a sphere in a power law fluid.
\newblock {\em J. Non-Newtonian Fluid Mech.}, 17:1--12, 1975.

\bibitem{Betan2015}
H.~Tabuteau, P.~Coussot, and J.~R. de~Bruyn.
\newblock Settling velocities of particulate systems part 17. {S}ettling
  velocities of individual spherical particles in power-law non-{N}ewtonian
  fluids.
\newblock {\em Int. J. Miner. Process.}, 143:125--130, 2015.

\bibitem{Ouyang2013}
L.~Ouyang, D.~Zhu, and A.~D. Hill.
\newblock Theoretical and numerical simulation of {H}erschel-{B}ulkley fluid
  flow in propped fractures.
\newblock In {\em Proceedings of the International Petroleum Technology
  Conference, 26-28 March, Beijing, China}, 2013.

\bibitem{Rich1954}
J.F. Richardson and W.N. Zaki.
\newblock Sedimentation and fluidization: Part i.
\newblock {\em Trans. Inst. Chem. Engrs.}, 32:35--47, 1954.

\bibitem{Gars1977}
J.~Garside and M.~R. Al-Dibouni.
\newblock Velocity-voidage relationships for fluidization and sedimentation in
  solid-liquid systems.
\newblock {\em Ind. Eng. Chem. Process Des. Dev.}, 16:206--214, 1977.

\end{thebibliography}


\end{document}